\let\clineorig\cline
\documentclass[pdflatex,sn-mathphys-num]{sn-jnl}
\let\cline\clineorig

\usepackage{graphicx}%
\usepackage{multirow}%
\usepackage{amsmath,amssymb,amsfonts}%
\usepackage{amsthm}%
\usepackage{mathrsfs}%
\usepackage[title]{appendix}%
\usepackage{xcolor}%
\usepackage{textcomp}%
\usepackage{manyfoot}%
\usepackage{booktabs}%
\usepackage{algorithm}%
\usepackage{algorithmicx}%
\usepackage{algpseudocode}%
\usepackage{listings}%
\usepackage{array}%

\DeclareMathOperator{\E}{\mathbb{E}}

\theoremstyle{thmstyleone}%
\newtheorem{theorem}{Theorem}[section]
\newtheorem{proposition}[theorem]{Proposition}%
\newtheorem{lemma}{Lemma}[section]

\theoremstyle{thmstyletwo}%

\theoremstyle{thmstylethree}%

\begin{document}

\title[A Black-Box Approach for Exogenous Replenishment in Online Resource Allocation]{A Black-Box Approach for Exogenous Replenishment in Online Resource Allocation}


\author[]{\fnm{Suho} \sur{Kang}}\email{suho\_kang@berkeley.edu}

\author[]{\fnm{Ziyang} \sur{Liu}}\email{ziyang\_liu@berkeley.edu}

\author[]{\fnm{Rajan} \sur{Udwani}}\email{rudwani@berkeley.edu}

\affil[]{\orgdiv{Department of Industrial Engineering and Operations Research}, \orgname{University of California, Berkeley}}



\abstract{In a typical online resource allocation problem, we start with a fixed inventory of resources and make online allocation decisions in response to resource requests that arrive sequentially over a finite horizon. 
We consider settings where the inventory is replenished over time according to an unknown exogenous process. 
We introduce black-box methods that extend any existing algorithm, originally designed without considering replenishment, into one that works with an arbitrary (adversarial or stochastic) replenishment process. Our approach preserves the original algorithm’s competitive ratio in regimes with large initial inventory, thereby enabling the seamless integration of exogenous replenishment into a large body of existing algorithmic results for both adversarial and stochastic arrival models.  }

\keywords{Online resource allocation, Exogenous replenishment, Competitive ratio}



\maketitle

\begingroup
\renewcommand{\thefootnote}{}
\footnotetext{An early version of this work was published in the 26th Conference on Integer Programming and Combinatorial Optimization \cite{kang2025black}.}
\endgroup

\section{Introduction}\label{SecIntro}

Online decision making problems where a fixed inventory of resources is allocated to sequentially arriving resource requests have been extensively studied, with applications in online advertising (\citet{mehta2007adwords}), airline pricing (\citet{talluri1998analysis}), personalized recommendation (\citet{golrezaei2014real}), sharing economy (\citet{rusmevichientong2020dynamic}) and many other domains. 
In most of these settings, the goal is to design an online algorithm that maximizes a global objective, such as total reward, over a possibly unknown planning horizon. A key theoretical goal is to find an algorithm with the highest possible \emph{competitive ratio}, which is the worst-case ratio of the expected total reward of an online algorithm to the expected total reward of the optimal offline solution (suitably defined for each problem). 
There is a vast body of literature on such algorithms for various settings, and we refer to \cite{mehta2013online, huangonline} for recent surveys. 

Almost all prior work assumes that the inventory is fixed initially and is not replenished during the planning horizon. In contrast, we are interested in settings where resources may be replenished through an (adversarial or stochastic) exogenous process. From this perspective, classical settings without replenishment are a special case where the replenishment is identically zero. The main question we try to answer in this paper is as follows.

\emph{Is there a way to generalize existing algorithms for adversarial or stochastic arrivals while preserving their performance guarantees in the presence of exogenous replenishment?}

\subsection{Previous Work} 
The essence of this question can be traced back to \citet{mehta2007adwords}, who introduced the classical Adwords problem for search ads. In their model, a fixed set of advertisers, each with a limited budget, is determined at the beginning of the planning horizon and opportunities for displaying ads arrive sequentially in an adversarial manner. They proposed the Inventory Balancing (IB) algorithm, which achieves a competitive ratio of $(1-1/e)$ in the \emph{large inventory regime}\footnote{The large budget/inventory assumption, in which the starting inventory of every resource is assumed to be much larger than the amount of resource that can be allocated to any individual request, is ubiquitous in the literature due to its practical relevance and technical simplicity. This regime is essentially equivalent to considering a fractional relaxation of the problem.}. 
As a practical extension of this model, they observed that IB's competitive ratio guarantees can be generalized to settings where new advertisers join the platform in an online manner and where budgets may be replenished during the planning horizon, provided \emph{each replenishment is sufficiently large}. Recent works \cite{vazirani2023towards, udwani2025adwords} study an extension of the Adwords problem where the (large) budget of each advertiser is initially unknown. The budget of an advertiser is revealed to the online algorithm immediately after it is exhausted and exhausted budgets are not replenished.

\citet{chan2009stochastic} introduced  the stochastic depletion problem which captures a wide range of online allocation problems, including the Adwords problem. They showed that a simple greedy (myopic) algorithm is $1/2-$competitive for adversarial arrivals and that this result holds even when the inventory is exogenously replenished with arbitrary (possibly small) amount of new inventory arriving in each period. 

While the IB algorithm has the highest achievable competitive ratio in the large inventory regime for many settings (including Adwords), its performance guarantee does not extend to scenarios where the replenishment process is arbitrary, and small amounts of inventory may arrive over time. Recently, \citet{feng2021robustness} provided the first generalization of IB, achieving a competitive ratio of $(1-1/e)$ in the large inventory regime under an arbitrary exogenous replenishment process. They employed a novel primal-dual analysis to overcome analytical challenges posed by the replenishment process. This approach works for the setting where both the arrival and replenishment processes are adversarial. Their analysis is also highly tailored to the IB algorithm and a specific problem formulation, and may not easily generalize. Even in the adversarial setting, there are many formulations in which IB is not the algorithm with the highest competitive ratio (see, for example, \cite{goyal2025asymptotically, ekbatani2024online, trobst2024almost}). 

Besides adversarial arrivals, other common settings include when requests arrive independently with known non-stationary probability distribution or independently with known or unknown stationary distribution, sometimes called online stochastic matching or Bayesian online matching (see, for example \cite{alaei2012prophet, mehta2013online, devanur2019near}). 
In this paper, we consider settings where the arrival process is stochastic and requests arrive independently according to a known, non-stationary probability distribution. To the best of our knowledge, no prior work has studied inventory replenishment in stochastic settings within the online resource allocation framework.

Our goal is to develop a black-box method that treats exogenous replenishment as a general feature rather than a setting-specific challenge, and to extend known algorithmic results from fixed-inventory (adversarial and stochastic) settings to those where the replenishment process may be either adversarial or stochastic.

\subsection{Our Contributions}


We consider exogenous replenishment within a generalization of the online resource allocation framework of \citet{devanur2019near}, which includes many well-studied settings. For the problems that are captured in this framework, we propose black-box methods that extend any online algorithm designed for settings without replenishment into one that retains the same performance guarantee with exogenous replenishment when the starting inventory is sufficiently large. 

Our black-box analysis extends the idea of ``batching'' introduced in Feng et al.~\cite{feng2021robustness}, and we consider both adversarial and stochastic settings for arrival processes as well as replenishment processes. In particular, when the arrival process is adversarial, our method transforms an instance with replenishment into an instance without replenishment and streams the transformed instance as input to the original online algorithm. In the transformation, we accumulate (batch) new inventory of a resource until it hits a certain threshold, after which we introduce it as a ``new'' type of resource that (one can imagine) was available since the beginning but incompatible with all prior arrivals. The rationale for this approach is to reduce an instance with arbitrary replenishment to one with only large replenishment. As observed in~\cite{mehta2007adwords}, a problem with only large replenishment can be reduced to an instance with no replenishment with the large starting inventory, since each replenishment can be treated as a new type of resource. Indeed, we prove that the proposed batching transformation is asymptotically lossless under the large starting inventory, regardless of the amount of replenishment in any period.

We then extend this idea to settings where the arrival process is stochastic. When the replenishment process is also stochastic (with a known, possibly non-stationary distribution), we replace each random replenishment with a deterministic value, treating the process as if it were fully predictable. Since this transformation can be performed before the algorithm begins, we can directly incorporate existing algorithms to effectively utilize the ``deterministic'' replenishment using the same batching idea. When the quantity of each replenishment is bounded above, we prove that this results in a lossless transformation.

The main technical challenge in this stochastic setting is that an action deemed feasible based on the pre-computed deterministic replenishment may become infeasible if the realized replenishment is insufficient. Our analysis, therefore, must carefully bound the probability of these events. We show that if each replenishment amount is small compared to the minimum starting inventory, the probability of such an event is asymptotically negligible.

On the other hand, we show that when a replenishment can be as large as the starting inventory, a lossless transformation is inherently impossible. This implies that when the arrival process is stochastic but the replenishment process is adversarial then a lossless extension is also impossible. Table 1 summarizes the outcomes across all four combinations of arrival and replenishment processes.

 
 

\emph{Overview of the analysis technique:} We analyze our black-box transformations in two steps: (i) we compare the performance of the transformed algorithm with the optimal offline solution to the transformed instance, and (ii) we compare the optimal offline solutions of the transformed instance and the original instance. In Step (i), we show that the competitive ratio is lower bounded by that of the original algorithm. In Step (ii), the analysis becomes independent of the online algorithm and depends solely on the properties of the general problem formulation. 
This two-step approach enables a simple and unified analysis of our approach. 

\begin{table}
\centering
\begin{tabular}{|cc|cc|}
\hline
\multicolumn{2}{|c|}{\multirow{2}{*}{}}                                      & \multicolumn{2}{c|}{\textbf{Arrivals (Request)}}      \\ \cline{3-4} 
\multicolumn{2}{|c|}{}                                                       & \multicolumn{1}{c|}{Adversarial} & Stochastic \\ \hline
\multicolumn{1}{|c|}{\multirow{2}{*}{\textbf{Exogenous Replenishment}}} & Adversarial & \multicolumn{1}{c|}{Yes}         & No        \\ \cline{2-4} 
\multicolumn{1}{|c|}{}                                         & Stochastic   & \multicolumn{1}{c|}{Yes}         & Yes$^*$ 
        \\ \hline
\end{tabular}
\vspace{0.1in}
\caption{Existence of Lossless Transformations in the large-inventory regime.
``Yes'' indicates that it is possible to transform any existing competitive
ratio result from the setting without replenishment to the setting with
exogenous replenishment. ``No'' indicates that such an extension is, in
general, impossible. ``Yes$^*$'' is conditional for the setting with stochastic
arrivals and replenishment, as the result depends on the maximum possible
amount of each replenishment.}
\label{tab:lossless_transformations}
\end{table}

\subsection{Outline}
In Section \ref{Notation}, we formally define a general framework for online resource allocation with exogenous replenishment. In Section \ref{Batch}, we introduce black-box algorithms applicable to problems with various arrival and replenishment processes. In Section \ref{Analysis}, we analyze these extensions, relying solely on the general problem formulation. We discuss further refinements of the black-box approach for more structured settings in Section~\ref{SecStructure}.
Finally, in Section \ref{Conclusion}, we conclude the paper by summarizing our findings and discussing open problems.

\section{A General Model with Stochastic Usage and Exogenous Replenishment}\label{Notation}


In this section, we formally present the online resource allocation framework of interest along with the relevant notation. We also illustrate the framework with several examples in Section \ref{AppendixExamplesCaptured}.

The framework has six main components: resources, (online) requests, actions, resource consumptions, exogenous replenishment, and rewards. In the following, we denote the set $\{1, \ldots, m\}$ by $[m]$, for any natural number $m$. 

\textbf{Resources.} Let \(\mathcal{I} = [n]\) represent the set of resource types, each resource type \(i \in \mathcal{I}\) having an initial inventory \(c_i\). Denote the minimum starting inventory among all resources as $c_{\min} = \min\limits_{i \in \mathcal{I}} c_i$.

\textbf{Requests.} Let \(\mathcal{J} = [m]\) denote the set of online requests. 
The sequence of requests is indexed from 1 (earliest) to \(m\) (latest). Let $Z$ denote the set of possible request types. When a request $j \in \mathcal{J}$ arrives, its request type $z \in Z$ is revealed to the online algorithm. The request type encapsulates a request's defining characteristics, such as resource requirements and rewards (which will be formally defined later).  

The set $Z$ is given to the online algorithm, but the prior knowledge of $\mathcal{J}$ depends on the arrival process. When the arrival process is stochastic, the total number of requests $m$ and the categorical distribution of request types are fully known in advance, prior to the arrival of the first request. Specifically, for each request \(j \in \mathcal{J}\), the online algorithm knows that the request type \(z \in Z\) will be drawn independently from the distribution \((p_{jz})_{z \in Z}\), where \(\sum_{z \in Z} p_{jz} = 1\). However, the type is revealed to the online algorithm when the request arrives. When the arrival process is adversarial, the online algorithm has no prior knowledge of the number of requests \(m\) or the distribution over request types. In this case, for simplicity, we assume that each request \(j\) is of type \(j\).

\textbf{Actions.} Let \(\mathcal{K} = [K]\) be the set of possible actions. On arrival of each request, we must select exactly one action from $\mathcal{K}$. 
The selection of actions is irrevocable.  Some actions may be infeasible if there is not enough inventory when the request arrives (discussed later). For illustration, in the online bipartite matching problem~\cite{karp1990optimal}, each action corresponds to either matching a request with a resource or ignoring the request. In the context of online assortment optimization~\cite{golrezaei2014real}, each action involves offering an assortment (set of resources) in response to a given request. 

\textbf{Resource Consumption.} 
If action $k \in \mathcal{K}$ is selected for request $j$ of type $z$, it consumes a random amount of (possibly multiple) resources. In the classical non-reusable resource setting (e.g. \cite{devanur2019near}), the amount of consumption is fixed once the action is chosen. Specifically, it requires \(A_{ijkz}\) units of resource type \(i \in \mathcal{I}\) after the action $k$ is chosen for the request. We let \(A_{i j k z}\) be a two-point random variable with support $\{0, b_{i j k z}\}$, where $b_{i j k z}$ is a deterministic quantity. The online algorithm observes its realization only after committing to action $k$ for request $j$ of type $z$.

Without loss of generality, we assume that $b_{ijkz} \leq 1$. This can be achieved by normalizing all resource consumption amounts and initial inventories by the maximum possible consumption for a single request. For instance, in the Adwords problem \cite{mehta2007adwords}, bids and budgets are divided by the largest possible bid to ensure this condition.

More generally, to capture settings with reusable resources where a resource is rented for a random duration and then returned (see Section 2.1 for a detailed explanation), we allow changes in resource consumption from previously chosen actions over time. Specifically, if action $k$ is selected for request $j$ of type $z$, it consumes \(A_{i j k z}(l)\) units of resource type \(i \in \mathcal{I}\) when request $l$ ($\geq j$) arrives. $A_{ijkz}(l)$ is a two-point random variable with support $\{0, b_{ijkz}(l)\}$, where $b_{ijkz}(l)$ is a deterministic quantity less than or equal to $1$. To ensure feasibility, we assume that resource consumption is non-increasing over time:
\[A_{i j k z}(l) \geq A_{i j k z}(l')\quad \forall l' > l,\, z \in Z.\] This generalizes non-reusable resource allocation where the consumption \(A_{i j k z}(l)\) is identical for all $l \ge j$.

The algorithm can choose action $k$ for request $j$ of request type $z$ only if the remaining inventory of each resource type $i \in \mathcal{I}$ is at least $b_{ijkz}(j)$. If $b_{ijkz}(j)$ is nonzero, we say that action $k$ for request $j$ of request type $z$ uses the resource type $i$. In other words, action $k$ is implementable only if the ``capacity constraints'' are satisfied for every resource used by the action. 

We also define $d$ as the maximum number of resource types used by any
single action. For example, $d=1$ in the online bipartite matching problem,
whereas in hypergraph matching, $d$ corresponds to the maximum cardinality
of a hyperedge.

We assume the existence of a trivial action $k_0 \in \mathcal{K}$, which requires no resources: $b_{i j k_0 z}(l) = 0$ for all $i \in \mathcal{I}$, $j \in \mathcal{J}$, $l \in \mathcal{J}$, $j \leq l$, and $z \in Z$. We also assume that resource consumption is independent across distinct arrivals and past decisions of the online algorithm. Specifically, we assume that $A_{i j k z}(l)$ is independent of $A_{i j' k' z'}(l)$ whenever $j \neq j'$, $k \neq k'$, or $z \neq z'$. Under these assumptions, for each request, we consider the sequence of diminishing resource consumption $\{A_{ijkz}(l)\}_{l \geq j, l \in \mathcal{J}}$, where the random variables in the sequence may be dependent due to monotonicity but are independent of past decisions and the resource consumption from other requests.

If the arrival process is stochastic, the full distribution of $\{A_{ijkz}(l)\}_{l \geq j, l \in \mathcal{J}}$ is known to the online algorithm at the beginning.
If the arrival process is adversarial, we assume that each request \(j\) is of type \(j\)
, and the distribution of $\{A_{ijkj}(l)\}_{l \geq j, l \in \mathcal{J}}$ remains unknown until the arrival of request $j$. 

\textbf{Exogenous Replenishment.} When request $j \in \mathcal{J}$ arrives, we also receive \(\zeta_i(j)\) new units for each resource type \(i \in \mathcal{I}\). The amount of replenishment is independent of the actions selected by the algorithm and the sequence of request types. Assuming that replenishment occurs upon the arrival of each request is without loss of generality, because decisions are only made at these moments. 

 When the exogenous replenishment process is adversarial, we have no prior knowledge of \(\zeta_i(j)\). Without loss of generality, we assume that \(\zeta_i(j)\) is a fixed, but unknown, non-negative constant with no additional assumptions beyond its non-negativity.
 
When the exogenous replenishment process is stochastic, the distribution of \(\zeta_i(j)\) is known at the beginning. The collection \(\{\zeta_i(j)\}_{i \in \mathcal{I}, j \in \mathcal{J}}\) is a family of independent random variables indexed over $\mathcal{I} \times \mathcal{J}$. 
For every $i$ and $j$, $\zeta_i(j)$ is a non-negative random variable
supported on $[0,M]$ with expected value $q_{ij}$, where $M\geq1$ is an upper bound on each stochastic replenishment. The realization of
$\zeta_i(j)$ is revealed upon the arrival of request $j$ and is independent
of previously chosen actions and resource consumption.%


\textbf{Rewards and Objective.} Upon the arrival of request $j \in \mathcal{J}$ of request type $z \in Z$, choosing an implementable action $k \in \mathcal{K}$ gives a random non-negative reward of $R_{ijkz}$ for each resource type $i \in \mathcal{I}$, where $R_{ijkz}$ and $R_{i'jkz}$ can be correlated for $i \neq i'$. The realization of reward $R_{ijkz}$ is independent of the replenishment process and prior decisions, and is revealed after we commit to an action $k$ for request $j$ of type $z$. For example, in the context of online assortment optimization, the random rewards capture the random selections made by customers. 

To capture the setting of multi-objective functions, we model the reward as proposed in Devanur et al. \cite{devanur2019near}, where each resource type $i$ corresponds to a different type of objective 
and the objective of the online algorithm is to maximize the minimum profit earned across all resource types, subject to the capacity constraints. Formally, for each $j \in \mathcal{J}$, let $z(j) \in Z$ denote its realized request type, and let $k(j) \in \mathcal{K}$ be the action chosen for request $j$ by the online algorithm. The profit associated with resource type $i$ is $\sum\limits_{j \in \mathcal{J}} R_{ijk(j) z(j)}$, and the objective function value of the algorithm becomes $\min\limits_{i \in \mathcal{I}} (\E [\sum\limits_{j \in \mathcal{J}} [R_{ijk(j) z(j)}])$. In Section \ref{AppendixExamplesCaptured}, we illustrate how this captures the reward maximization objective used in many formulations. 

If the arrival process is stochastic, the distribution of \( R_{ijkz} \) is known a priori.
If the arrival process is adversarial, we assume each request $j$ is of type $j$, and the distribution of reward $R_{ijkj}$ is revealed upon the arrival of request \( j \).  



\textbf{Offline Benchmark and Competitive Ratio.} Throughout this paper, we focus on worst-case (competitive ratio) analysis over a family of instances $\mathcal{H}$. We compare online algorithms with a clairvoyant algorithm, OPT, which has full knowledge of the sequence of requests $j \in \mathcal{J}$ (and request types $z(j) \in Z$), the set of feasible actions $\mathcal{K}$, and the realizations of the replenishment process $\{\zeta_i (j)\}_{i\in \mathcal{I}, j\in \mathcal{J}}$ before the first request arrives. However, while OPT has prior knowledge of the distributions of resource consumption $\{A_{ijkz}(l)\}_{l \geq j, l \in \mathcal{J}}$'s and rewards $R_{ijkz}$'s, it does not know their actual realizations until the arrival of request $l>j$ for consumption or until action $k$ is chosen for rewards, just like any other online algorithm.\footnote{No interesting competitive ratio results are possible against the benchmark that knows the realization of $A_{ijkz}(l)$'s in advance. See \cite{goyal2023online} for an example.}

Let $\text{ALG}$ denote an online algorithm for a family of instances $\mathcal{H}$. For each instance $H \in \mathcal{H}$, let $\text{ALG}(H)$ and $\text{OPT}(H)$ denote the expected objective function value achieved by ALG and OPT, respectively. Then, the competitive ratio of ALG for $\mathcal{H}$, is defined as follows:
\begin{equation*}
    \text{Competitive Ratio of ALG for $\mathcal{H}$ := } \min_{H \in \mathcal{H}} \frac{\text{ALG}(H)}{\text{OPT}(H)}.
\end{equation*}

We note that the competitive ratio provides a common
performance measure across the adversarial and stochastic, possibly non-stationary, arrival and replenishment models considered in our
framework. In contrast, sublinear additive regret against our clairvoyant
offline benchmark is not attainable in general, as illustrated by the construction in Theorem~\ref{thm4.3} later.

Nevertheless, in more structured online resource allocation
settings, such as certain settings with i.i.d.\ arrivals, algorithms can
achieve asymptotically optimal competitive ratios approaching $1$, and regret
provides a finer measure of their performance
\citep{vera2021bayesian,li2022online}.
Investigating whether black-box transformations can preserve
such low-regret guarantees under exogenous replenishment is an
interesting direction for future research, but is beyond the
scope of this paper.


\subsection{Examples of Online Resource Allocation Problems}\label{AppendixExamplesCaptured}
To understand the different components of our framework, we will discuss several well-studied settings it captures. First, we'll introduce some basic notation used throughout this paper.

We employ the indicator function, denoted by $\mathbb{I}(\cdot)$, which equals $1$ if its argument is true and $0$ otherwise. Additionally, to describe the asymptotic behavior of functions, especially when analyzing competitive ratios in the large-inventory regime, we employ the following family of Landau notations:
\begin{enumerate}[\indent (i)] 
    \item $f(c) = o(g(c))$ means $\lim_{c \rightarrow \infty} \frac{f(c)}{g(c)} = 0$.
    \item $f(c) = \mathcal{O}(g(c))$ means $\limsup_{c \rightarrow \infty} \left|\frac{f(c)}{g(c)}\right| < \infty$.
    \item $f(c) = \Omega(g(c))$ means $\liminf_{c \rightarrow \infty} \left|\frac{f(c)}{g(c)}\right| > 0$.
\end{enumerate}
Now, we are ready to discuss the examples that fall into our framework.

\begin{itemize}
    \item \textbf{$b-$matching problem and Adwords.} Generalizing the classical online bipartite matching problem \cite{karp1990optimal}, the b-matching problem \cite{kalyanasundaram2000optimal} involves matching incoming requests to a fixed set of resources, each with a limited capacity (or ``budget" represented by $b_i$ for $i \in \mathcal{I}$). Every online request is matched to at most one resource and consumes a unit amount of it. To capture this within our framework, let the action set $\mathcal{K}= [n] \cup \{k_0\}$, where action $k \in [n]$ represents matching a request to resource type $k$, and action $k_0$ corresponds to ignoring the request. Each request type $z \in \mathcal{Z}$ is characterized by a set of compatible resources, denoted as $\mathcal{I}(z) \subseteq \mathcal{I}$. Resources are non-reusable and resource consumption is deterministic. Specifically, if action $k \in \mathcal{I}(z)$ is chosen for a request of type $z$, the consumption for resource type $i$ at time $l \ge j$ is simply 
    $A_{ijkz}(l) = \mathbb{I}\left(k \in \mathcal{I}(z)\right) \mathbb{I}\left(i =k \right)$. 
    Matching an arrival of type $z$ to resource $i$ generates reward $w_i$ when $i\in \mathcal{I}(z)$, and we capture this by setting $R_{ijkz} = w_k \mathbb{I}\left(k \in \mathcal{I}(z)\right)\, \forall i\in \mathcal{I}$.

    Similarly, to capture the Adwords problem \cite{mehta2007adwords}—where advertisers ($\mathcal{I}$) bid on online queries ($\mathcal{J}$)— for each request of type $z \in \mathcal{Z}$ we set $A_{ijkz}(l) = b_{iz}\mathbb{I}\left( i=k \right)$, where $b_{iz}$ represents the amount bid by advertiser (resource) $i$ on query (request) of type $z$. 
   We also let $R_{ijkz} = b_{iz}\, \forall i\in \mathcal{I}$ and $k \neq k_0$, and $R_{ijk_0z} = 0$.

    \item \textbf{Stochastic rewards and assortment optimization.} 
    In the setting of online bipartite matching with stochastic rewards, every online request is matched to at most one resource and a proposed match may, with some probability, be rejected by the request. When the match is rejected, the resource is not used, there is no reward, and the request is not re-matched. This setting was introduced by Mehta and Panigrahi \cite{mehta2012online} and further generalized by Mehta et al. \cite{mehta2014online}. The problem of online assortment optimization, introduced by Golrezaei et al. \cite{golrezaei2014real}, is another generalization of matching with stochastic rewards. In this problem, each arrival is offered a set of options (an assortment) and the arrival chooses at most one resource from the offered assortment. The probability of a resource being chosen depends on the offered assortment. 

    For the online matching problem with stochastic rewards, we can use the same setting as the $b$-matching problem, but with different resource consumption $A_{ijkz}(l)$'s and rewards $R_{ijkz}$'s. For each request $j$ of type $z$, we draw an independent Bernoulli random variable $(X_{ijz})_{i \in \mathcal{I}}$ based on the rejection probabilities, and define $A_{ijkz}(l) = (1-X_{ijz})\mathbb{I}\left(k \in \mathcal{I}(z)\right) \mathbb{I}\left(i =k \right)$, and $R_{ijkz} = (1-X_{ijz})w_k \mathbb{I}\left(k \in \mathcal{I}(z)\right)$.
        For the assortment optimization problem, we define the action set $\mathcal{K}$ as the set of all possible assortments (e.g., all subsets of $\mathcal{I}$) and specify the random variables $(X_{ijz})_{i \in \mathcal{I}}$ based on the choice model of customer type $z$.

    \item \textbf{Reusable resources.} A recent line of work considers settings with reusable resources. These reusable resources are allocated for some duration and then become available again for re-allocation after this duration (see, for example, \cite{rusmevichientong2020dynamic},  \cite{gong2022online}, \cite{goyal2025asymptotically}, \cite{feng2021robustness}, \cite{feng2024near}, and \cite{ekbatani2024online}). The timing of a resource's next availability may depend on the specific request or the time it was matched for the last time.
    
Our framework can incorporate such non-stationary reusability for each resource through the time-dependent resource consumption parameter, $A_{ijkz}(l)$. Specifically, our model allows for an arrival-dependent usage distribution: if a resource of type $i$ is used via action $k$ for request $j$ of type $z$, we consider whether it remains consumed or becomes available again at subsequent request arrivals. If the resource becomes available when a later request $l > j$ arrives, this corresponds to the realization $A_{ijkz}(l) = 0$, indicating it is no longer being consumed by request $j$ at time $l$. Conversely, if $A_{ijkz}(l) > 0$, the resource is still considered in use by request $j$.

    \item \textbf{Network revenue management and Hypergraph Matching.} 
    In the hypergraph matching problem, requests arrive for bundles of resources, and the entire bundle becomes infeasible if any resource in a bundle is unavailable. In network revenue management, bundles may have different prices, and the goal is to maximize total revenue. We refer to Talluri and Van Ryzin \cite{talluri1998analysis} and Buchbinder and Naor \cite{buchbinder2009online} for classic results on these problems. A sample of more recent work includes Ma et al. \cite{ma2020approximation}, Trobst and Udwani \cite{trobst2024almost}, Jiang \cite{jiang2023constant}, Ma et al. \cite{ma2024online}. 

To incorporate these problems into our framework, we let $\mathcal{K}$ be the set of bundles, where each bundle $k$ uses a specified $b_{ik}$ amount of resource type $i$. Each request (customer) type $z \in Z$ is characterized by a subset of feasible bundles $K(z)\subseteq \mathcal{K}$. If bundle $k$ has a price $w_{jkz}$ for request $j$ of type $z$, we let $R_{ijkz}=w_{jkz}\,\forall i\in \mathcal{I}$. 

\end{itemize} 

\subsection{Main Results}


To describe our main results, we introduce some notation. Given a family of instances $\mathcal{H}$ from our online resource allocation framework, consider the subset $\mathcal{H}_0 \subset \mathcal{H}$ that consists of instances without exogenous replenishment, that is, $\zeta_i(j) = 0$ for all $i \in \mathcal{I}$ and $j \in \mathcal{J}$. We assume that $\mathcal{H}_0$ is closed in the following sense: For every instance in $\mathcal{H}_0$, adding or removing a resource type results in a modified instance that remains in $\mathcal{H}_0$. For example, given an instance of the online bipartite matching problem, adding or removing a specific resource type still produces an instance of the online bipartite matching problem. Depending on the choice of $\mathcal{H}$, $\mathcal{H}_0$ may correspond to a well-studied problem setting. The assumption on $\mathcal{H}_0$ is innocuous, and it is not hard to see that it holds for all the 
settings discussed in Section \ref{AppendixExamplesCaptured}. 

For any $c>0$, let $\mathcal{H}(c) \subset \mathcal{H}$ and
$\mathcal{H}_0(c) \subset \mathcal{H}_0$ denote the subsets of instances
whose minimum initial inventory satisfies $c_{\min} \geq c$.
Here, $c_{\min}$ denotes the minimum initial inventory of a particular
instance, whereas $c$ is a lower-bound parameter used to define the
corresponding family of instances. Given an algorithm ALG operating on instances in $\mathcal{H}_0(c)$, let $\text{ALG}(H)$ represent the expected objective function value achieved by ALG on $H \in \mathcal{H}_0(c)$. Define $\alpha(c)$ as the parametric competitive ratio of ALG over
$\mathcal{H}_0(c)$:


\begin{equation*}
    \alpha(c) := \min_{H \in \mathcal{H}_0 (c)} \frac{\text{ALG}(H)}{\text{OPT}(H)}.
\end{equation*}
For example, $\alpha(c)=\left(1-\frac{1}{e}\right)
-\frac{\mathcal{O}(1)}{c}$ for the \textsc{Balance} algorithm for online
$b$-matching \cite{kalyanasundaram2000optimal}. As an example under stochastic arrivals, Alaei et
al.~\cite{alaei2012prophet} give a randomized algorithm for the Adwords
problem with parametric competitive ratio
$\alpha(c)=1-\mathcal{O}(1/\sqrt{c})$.

Without loss of generality, we assume that ALG will not deliberately consume resources if it gains no reward in return. That is, for each request $j$ of type $z$, ALG selects either a trivial action $k_0$ or a non-trivial action $k \in \mathcal{K}$ with a non-zero reward $R_{ijkz}$ for at least one resource type $i \in \mathcal{I}$. If, counter to this, an algorithm were to select an action that consumes resources without gaining any rewards, we could simply replace that selected action with the trivial action $k_0$ without decreasing the overall objective function value. Also, we restrict attention to algorithms whose decisions are unaffected by adding resource types whose associated actions have zero reward for all requests realized so far. This mild condition is satisfied by the standard (optimally competitive) algorithms considered in the online resource allocation literature and allows future exogenous replenishment to be represented as new resource types without altering earlier decisions in the analysis.


Recall that $d$ denotes the maximum number of resource types used by any single action. Also, recall that each stochastic replenishment $\zeta_i(j)$ for any resource type $i$ and request $j$ is bounded above by $M$, which determines the range of uncertainty in replenishment. Theorems \ref{thm4.1} and \ref{thm4.2} state the theoretical guarantees obtained through our black-box extensions. The corresponding extensions are described in Section \ref{Batch}, and the detailed proofs of Theorems \ref{thm4.1} and \ref{thm4.2} are provided in Section \ref{Analysis}.

\begin{theorem}\label{thm4.1}

     Given a family of instances $\mathcal{H}$ with exogenous replenishment and an adversarial arrival process, suppose that ALG achieves a competitive ratio of at least $\alpha(c)$ for $\mathcal{H}_0(c)$. There exists an algorithm, ALG-B, that achieves a competitive ratio of at least $\alpha\left(\sqrt{c}\right)\left(1-\mathcal{O}\left(\sqrt{\frac{\log(cd)}{c}}\right)\right)$ for $\mathcal{H}(c)$.

\end{theorem}

\begin{theorem}\label{thm4.2}
Given a family of instances $\mathcal{H}$ and $c>0$ under a stochastic arrival process and a stochastic replenishment process, suppose that $\text{ALG}$ achieves a competitive ratio of at least $\alpha(c)$ for the set of instances $\mathcal{H}_{0}(c)$. 
There exists an algorithm, ALG-B, that achieves a competitive ratio of at least $\alpha\left(
\mathcal{O}\left(
M^{\frac{1}{3}} c^{\frac{2}{3}} \log^{\frac{1}{3}}(cd)
\right)\right)
\left(
1-\mathcal{O}\left(
M^{\frac{1}{3}} c^{-\frac{1}{3}} \log^{\frac{1}{3}}(cd)
\right)
\right).$ 
\end{theorem}

In the large inventory regime (\(c \to +\infty\)), Theorem \ref{thm4.1} implies that there exists an algorithm with asymptotic competitive ratio equal to $\lim_{c\to +\infty} \alpha(c)$, which means that it is possible to losslessly transform an algorithm from a setting with fixed inventory into an algorithm for the corresponding generalization of the setting with exogenous replenishment. 
Furthermore, Theorem \ref{thm4.2} demonstrates that we can losslessly transform an algorithm designed for fixed-inventory settings into one capable of handling a stochastic replenishment process when
$M=\mathcal{O}(c^{1-\nu})$ for some constant $\nu>0$.
We summarize the competitive ratio achievable by ALG-B in some specific cases in Table~\ref{T1}.\footnote{
Directly applying Theorem~\ref{thm4.1} to assortment optimization under
adversarial replenishment yields the guarantee
\[
\left(1-\frac{1}{e}\right)
\left(
1-\mathcal{O}\left(
\sqrt{\frac{\log(nc)}{c}}
\right)
\right).
\]
However, under the standard substitutability assumption, since each realized purchase consumes at most one resource
type, a tailored analysis avoids the union bound over all products in
the offered assortment and thus removes the dependence on $n$ in the guarantee. See Section~\ref{SecStructure} for
details. This shows that
the guarantee obtained by our general black-box analysis can be improved in
more structured settings.
} The results are based on a general method to construct and analyze a new algorithm (ALG-B) based on an existing algorithm (ALG), which we refer to as the ``batching extension''.

\begin{table}[h!]
    \centering
    \begin{tabular}{|c|c|c|}
    \hline
    \small{\textbf{Problems}}
    & \small{\textbf{Adversarial}}
    & \small{\textbf{Stochastic}} \\
    \hline

    Adwords \cite{mehta2007adwords,alaei2012prophet,devanur2019near}
    &
    $\left(1-\frac{1}{e}\right)
    \left(
    1-\mathcal{O}\left(
    \sqrt{\frac{\log(c)}{c}}
    \right)
    \right)$
    &
    $1-\mathcal{O}\left(
    \left(\frac{M}{c}\log(c)\right)^{\frac{1}{3}}
    \right)$
    \\ \hline

    Assortment Optimization
    \cite{golrezaei2014real,devanur2019near}
    &
    $\left(1-\frac{1}{e}\right)
    \left(
    1-\mathcal{O}\left(
    \sqrt{\frac{\log(c)}{c}}
    \right)
    \right)$
    &
    $1-\mathcal{O}\left(
    \left(\frac{M}{c}\log(nc)\right)^{\frac{1}{3}}
    \right)$
    \\ \hline

    Hypergraph Matching
    \cite{trobst2024almost,devanur2019near}
    &
    $\frac{1}{d+1}
    \left(
    1-\mathcal{O}\left(
    \sqrt{\frac{\log(cd)}{c}}
    \right)
    \right)$
    &
    $1-\mathcal{O}\left(
    \left(\frac{M}{c}\log(cd)\right)^{\frac{1}{3}}
    \right)$
    \\ \hline

    Online Matching of Reusable
    & & \\
    Resources (Stochastic Durations)
    \cite{feng2024near, goyal2025asymptotically}
    &
    $\left(1-\frac{1}{e}\right)
    \left(
    1-\mathcal{O}\left(
    \sqrt{\frac{\log(c)}{c}}
    \right)
    \right)$
    &
    $1-\mathcal{O}\left(
    \left(\frac{M}{c}\log(c)\right)^{\frac{1}{3}}
    \right)$
    \\
    \hline
    \end{tabular}

    \vspace{0.1in}
    \caption{Lower bound on the competitive ratio of the batching algorithm
    in the large-inventory regime (under adversarial or stochastic arrival processes).}
    \label{T1}
\end{table}

Furthermore, when $M = \Omega(c)$, we demonstrate that a lossless transformation is, in general, impossible, as stated in Theorem \ref{thm4.3}. The high-level reason for this 
result is that in settings with stochastic arrivals, online algorithms can often leverage distributional knowledge of the arrival process to achieve performance similar to offline but a single large individual replenishment can significantly change the offline solution, making it impossible for an online algorithm to achieve the same performance as offline.
The detailed proof is also provided in Section \ref{Analysis}. 

\begin{theorem}\label{thm4.3}
Fix any constant $\gamma\in(0,1]$. There exists a family of instances
$\mathcal{H}$ under a stochastic arrival process and a stochastic
replenishment process with $M=\gamma c$ such that the corresponding
setting without exogenous replenishment admits an online algorithm with
competitive ratio $1$, whereas every online algorithm for $\mathcal{H}$
has competitive ratio at most
\[
1-\frac{\gamma}{2+3\gamma}.
\]
\end{theorem}


As a corollary of Theorem \ref{thm4.3}, we have that a lossless transformation is impossible when the arrival process is stochastic but the replenishment process is adversarial, completing Table~\ref{tab:lossless_transformations} in Section~\ref{SecIntro}.

\section{Black-Box Algorithmic Extensions}\label{Batch}


Consider a set of instances $\mathcal{H}_0 \subset \mathcal{H}$, where no exogenous replenishment occurs in $\mathcal{H}_0$. Given an algorithm ALG designed for $\mathcal{H}_0$, our goal is to construct a black-box extension, denoted by $\text{ALG-B}$, that can handle any instance $H \in \mathcal{H}$ and retains the same asymptotic performance guarantee in the large-inventory regime.

In particular, based on the replenishment in $H$, we are interested in constructing a ``batched instance'' $H_B \in \mathcal{H}_0$ in an online manner and streaming this instance into ALG. The resulting online algorithm (for $H$) is denoted by ALG-B. The core idea behind creating $H_B$ is simple: (i) wait until the cumulative exogenous replenishment for a resource type reaches a certain threshold $c_B$, and (ii) treat the accumulated/batched replenishment as a new resource type and make allocation decisions guided by ALG with an augmented set of resource types. In this paper, we call such a black-box algorithmic extension a ``batching extension'' of an algorithm and discuss these steps in more detail in the following subsections.

In Section \ref{sBatchingAd}, we discuss the batching extension under an adversarial arrival process. In this case, we define the batched replenishment process assuming that the replenishment process is adversarial. This assumption is without loss of generality because we can always choose not to use prior knowledge, and we can treat the stochastic replenishment process as if it were an adversarial replenishment process.

Then, in Section \ref{sBatchingSto}, we define the batching extension under a stochastic arrival process. Due to the impossibility result of Theorem \ref{thm4.3}, we only consider the setting with a stochastic replenishment process.

\subsection{Batching Extension under an Adversarial Arrival Process}\label{sBatchingAd}




Recall an observation from \cite{mehta2007adwords}: when the amount of each replenishment is large, we can incorporate each replenishment as a new resource type, maintaining the same asymptotic competitive ratio.  One way to leverage this observation for an arbitrary replenishment process is to wait until the cumulative replenishment of a resource is large enough (compared to initial inventory). This is the idea behind the batching threshold $c_B$. If this threshold is too small, the resulting new resource types may begin with a small amount of inventory, causing the transformed instance to no longer be in the large inventory regime. Indeed, the size of the batch will determine the minimum inventory in the transformed instance. 
On the other hand, choosing a large value of $c_B$ may lead to a considerable amount of loss in reward compared to offline, since the online algorithm will not utilize newly available inventory for a resource until the cumulative replenishment exceeds the threshold, whereas the optimal offline solution can use new inventory immediately. 

In summary, choosing an appropriate threshold $c_B$ is critical for the batching extension. A larger value of $c_B$ allows us to leverage existing results in the large-inventory regime, even when the replenishment process is adversarial. Conversely, a smaller value of $c_B$ reduces excessive delays caused by waiting for batching. We balance this tradeoff by setting $c_B= \sqrt{c_{\min}}$ (recall that $c_{\min} = \min\limits_{i \in \mathcal{I}} c_i$) under adversarial arrival and replenishment processes, and we show in  Section \ref{Analysis} that this gives the best theoretical guarantee in the context of our analysis. 


\subsubsection{Batched Replenishment Process under an Adversarial Replenishment Process}\label{3.1.1.}
Recall that under an adversarial arrival process we omit request type $z$ from the notation. Upon the arrival of request $j$, we observe $\zeta_i(j)$ amount of replenishment for resource $i$. Based on the choice of $c_B$, we formally define a batched replenishment process $\left\{\zeta_i^B(j)\right\}_{i \in \mathcal{I}, j \in \mathcal{J}}$ 
as follows. For each resource type $i$, initialize $\zeta^B_i(0) = 0$ and $\bar{c}_i = 0$. The variable $\bar{c}_i$ will capture the new inventory of $i$ that has been received but has not been batched to use in the online algorithm. On arrival of a request $j$, we update $\bar{c}_i = \sum_{l \in [j]} \zeta_i(l) - \sum_{l \in [j-1]} \zeta^B_i(l)$. If $\bar{c}_i \ge c_B = \sqrt{c_{\min}}$, then we set $\zeta^B_i(j) = \bar{c}_i$, otherwise we let $\zeta^B_i(j)$ stay equal to 0. Notice that $\zeta^B_{i} (j)$ is either $0$ or at least $c_B$. In the batched replenishment process, on arrival of request $j$, we receive $\zeta^B_{i} (j)$ units of resource type $i$ instead of $\zeta_i(j)$. 

By definition, we have the following relation:
\[\sum_{l \in [j]} \zeta_i(l) - c_B \leq \sum_{l \in [j]} \zeta^B_i(l) \leq \sum_{l \in [j]} \zeta_i(l), \quad \forall j \in \mathcal{J}.\]
Therefore, upon each arrival, we can transform a general instance $H$ with replenishment process $\{ \zeta_i(j) \}_{i \in \mathcal{I}, j \in \mathcal{J}}$ into an instance $H_B$ with replenishment process $\{ \zeta_i^B(j) \}_{i \in \mathcal{I}, j \in \mathcal{J}}$ in an online manner. Importantly, every non-zero replenishment in $H_B$ is large.

\subsubsection{Batched Instance and Batching Extension}\label{3.1.2.}

The high-level idea of the batching extension (algorithm) is as follows. Based on the batched replenishment process $\{\zeta_i^B(j)\}_{i \in \mathcal{I}, j \in \mathcal{J}}$ that can be constructed in an online manner, for each $\zeta_i^B(j) > 0$ (which implies $\zeta_i^B(j) \geq c_B$), we introduce a new resource type $i_j$ with the starting inventory $\zeta_i^B(j)$. This new resource type is then treated as if it had existed from the beginning. By replacing all nonzero batched replenishment with such new resource types, the transformed instance effectively becomes an instance in $\mathcal{H}_0$ with minimum starting inventory at least $c_B$. This transformed instance is then provided as input to the existing algorithm ALG to determine the next allocation decisions. This batching extension is denoted as $\text{ALG-B}$. See Appendix \ref{appendix-ex} for an illustrative example.

It is worth noting that ALG-B remains an online algorithm. Note that an online algorithm ALG, designed for fixed-inventory settings, determines its action for a given request based on the sequence of chosen actions and the observed filtration (i.e., the information revealed up to that point), including details of requests, available resources, and feasible actions. ALG-B operates by dynamically transforming the original instance $H$ in an online manner, creating a modified instance description that is continuously fed as input to ALG. Specifically, as ALG-B introduces new batched resource types (e.g., $i_j$) and their associated actions, the set of available resources and actions within the instance description can also be updated in an online manner. This transformation ensures that ALG-B operates as a well-defined online algorithm, because its decisions are always based solely on information observed up to the current moment, with future modifications to the instance not affecting past or present choices.

Having clarified the conceptual workings and online properties of the batching extension, we now elaborate on its details. On the arrival of request $j$, before choosing an action for request $j$, we generate copies of each action $k \in \mathcal{K}$ by replacing the original resources in $\mathcal{I}$ with their identical counterparts, as additional batched resource types may now be available. Formally, for each request $j \in \mathcal{J}$, we update the available set of actions $\mathcal{K}(j)$ recursively. First, initialize $\mathcal{K}(0) := \mathcal{K}$ and assume we have $\mathcal{K}(j-1)$. When request $j$ arrives, we construct $\mathcal{K}(j)$ as follows: set $\mathcal{K}_0(j) := \mathcal{K}(j-1)$, and for each $i \in \mathcal{I}$, define $\mathcal{K}_i(j) := \mathcal{K}_{i-1}(j)$ and check if a new batched resource type $i_j$ is created. If so, for each action $k \in \mathcal{K}_{i-1}(j)$, we create a duplicate action $k(i_j)$ by replacing the usage of resource type $i$ with resource type $i_j$ and add it to $\mathcal{K}_{i}(j)$ with the same reward. After checking all resource types, set $\mathcal{K}(j) := \mathcal{K}_n(j)$. Based on the current sequence of chosen actions, instance information, and available actions $\mathcal{K}(j)$, we use ALG to select an action as if batched resource types and actions in $\mathcal{K}(j)$ were present from the start. Note that if ALG selects an action $k'$ that operates on a newly introduced batched resource type (i.e., $k' \notin \mathcal{K}$), this choice is then implemented by executing the corresponding original action $k \in \mathcal{K}$ (from which $k'$ was duplicated) on the original instance $H$, with resource consumption and reward coupled as defined for action $k$.

Although the identical resource $i_j$ is introduced later, we can treat $i_j$ and its associated actions in $\mathcal{K}(j)$ as if they were present from the beginning by controlling rewards. Assume that we already know all batched resource types and available actions $\mathcal{K}(m)$ at the start of the algorithm. If an action $k \in \mathcal{K}(m)$ has not been introduced for the current request $j$ (i.e., $k \in \mathcal{K}(m) \setminus \mathcal{K}(j)$), we set its reward to 0. Conversely, if an action $k$ is activated when request $j$ arrives (i.e., $k \in \mathcal{K}(j)$), selecting an action $k$ for request $j$ yields the original reward. In other words, we obtain a nonzero reward only if the action uses resource types defined at or before the arrival of request $j$. Recall that the algorithm selects either a trivial action $k_0$ or a non-trivial action $k \in \mathcal{K}(m)$ with a non-zero reward $R_{ijkj}$ for some $i \in \mathcal{I}$ and each request $j \in \mathcal{J}$. Although we create the identical resource \(i_j\) later, by the assumptions underlying ALG, the output (i.e., the sequence of chosen actions) of the algorithm remains consistent up to request \(j\), as if we generated the resource type \(i_j\) and its associated actions from the beginning. The formal definition of ALG-B is presented in Algorithm \ref{alg:ALG-B}.

It is worth noting that the execution of ALG remains
consistent in the batched instance. Consider a resource copy $i_j$
introduced at request $j$. In the batched instance $H_B$, the actions
associated with $i_j$ have zero reward for every request $\ell<j$.
Recall that, by the condition stated in Section~2.2, the presence of such resource types does not affect the decisions of ALG before request $j$. Therefore, the execution in which $i_j$ is introduced at request $j$ can be coupled with the execution in which $i_j$ is present from the beginning, so that they agree through request $j$. Repeating this argument for each newly introduced resource copy shows that the execution of ALG-B is consistent with the execution of ALG on $H_B$.

After updating the set of actions $\mathcal{K}(m)$ and resource types $\mathcal{I}$ upon the arrival of the last request $m$, we obtain a fully transformed instance $H_B$ without exogenous replenishment, which we call the batched instance. Note that although we do not know the complete $H_B$ in advance, $\text{ALG-B}(H)$ makes allocation decisions that are consistent with those made by $\text{ALG}(H_B)$.

\begin{algorithm}[h!]
\caption{Definition of $\text{ALG-B}$ under an adversarial replenishment process}\label{alg:ALG-B}
\begin{algorithmic}
\State \textbf{Input:} A set of resource types $\mathcal{I}$, the initial inventory $c_i$ for each $i \in \mathcal{I}$, \\ \ \ \ \ \ \ \ \ \ \ and a set of actions $\mathcal{K}$.
\State{Set $c_{\min} = \min_{i \in \mathcal{I}} c_i$, and $c_B := \sqrt{c_{\min}}$.} \Comment{Batching threshold}
\State {Initialize $K_{past} = [\quad]$.} \Comment{Sequence of actions chosen by ALG}
\State {Initialize $\mathcal{K}(0) := \mathcal{K}$.} \Comment{Set of actions}
\For {each resource type $i \in \mathcal{I}$}
    \State {Initialize $\bar{c_i} := 0 $}  \Comment{Cumulative replenishment for batched replenishment}
\EndFor
\For {each request $j \in \mathcal{J}$} \Comment{Process each arrival online}
    \State {Initialize $K_0(j) := \mathcal{K}(j-1)$.}
    \For {each resource type $i \in \mathcal{I}$}
    \State $\bar{c_i} := \bar{c_i} + \zeta_i(j)$. \Comment{Keep track of cumulative replenishment}
    \State $\mathcal{K}_i(j) := \mathcal{K}_{i-1}(j)$. \Comment{Keep track of available actions}
        \If{$\bar{c_i} \geq c_B$} \Comment{Cumulative replenishment exceeds threshold}
            \State $\mathcal{I} := \mathcal{I} \cup \{i_j\}$. \Comment{Create an identical resource type}
            \State $c_{i_j} := \bar{c}_i$.
            \State $\mathcal{K}_i(j) := \mathcal{K}_i(j) \cup \{k(i_j): k \in \mathcal{K}_i(j)\}$. \Comment{Add copies of actions}
            \State $\bar{c_i} := 0$. \Comment{Reset cumulative replenishment}
        \EndIf
    \EndFor
    \State $\mathcal{K}(j) := \mathcal{K}_n(j)$. \Comment{Define a set of available actions for request $j$}
    \State{Choose an action $\kappa(j)$ using ALG on the instance with $\mathcal{I}$ and $\mathcal{K}(j)$, based on $K_{past}$ and the observed filtration.}
    \State{Append $\kappa(j)$ to $K_{past}.$
    \State{Implement the original action in $\mathcal{K}$ corresponding to $\kappa(j)$.}}
\EndFor
\end{algorithmic}
\end{algorithm}

\subsection{Batching Extension under a Stochastic Arrival Process}\label{sBatchingSto}


Under a stochastic arrival process, the probability distributions of resource consumption and rewards are known in advance, and ALG may leverage this prior information (e.g., see \cite{devanur2019near}). Therefore, under a stochastic replenishment process\footnote{Recall that we do not consider an adversarial replenishment process in this section due to the impossibility result of Theorem \ref{thm4.3}.}, to leverage the prior distribution of the replenishment process, 
we replace the random replenishment process with a deterministic fluid counterpart. Then, we construct a batched replenishment process, analogous to the one used in the adversarial case, for this fluid replenishment. Note that this batched fluid replenishment process can be computed in advance, purely based on the prior information of the replenishment process.

At a high level, similar to the adversarial case, we convert each batched replenishment into  a new resource type. This transforms the instance $H$ (with exogenous replenishment) into a batched instance $H_B \in \mathcal{H}_0$ (with fixed capacity). Then, we would like to run the original algorithm ALG on the batched instance and implement the original corresponding action if possible.

The main challenge in the stochastic setting, which does not arise in the adversarial case, is that the action by ALG may not be implementable due to the potential deviation between the actual (random) replenishment realizations and the predefined (deterministic) batched fluid replenishment. Note that in the adversarial setting, there is no such deviation since batched replenishment is derived directly from observed amounts in an online manner.

To overcome this difficulty, we run ALG in a virtual environment that is decoupled from the realized replenishment process. Also, when defining the fluid replenishment process for ALG, we slightly underestimate the expected amount of replenishment. This ensures that there is sufficient inventory to select and the action selected by ALG (executed on the fluid process) with high probability. Nonetheless, capacity violation may still occur when the action selected by ALG on the batched (fluid) instance $H_B$ is not implementable and in this case we replace the action with the trivial action $k_0$. However, ALG is oblivious to such an event and we continue to pretend that the action it selected has been successfully implemented. In short, we (i) Compute the batched (fluid) instance $H_B$ in advance (without observing any realizations). (ii) Pretend to ALG that the realized instance is always $H_B$. (iii) When the action selected by ALG is unimplementable, we select the trivial action $k_0$.

\subsubsection{Fluid Replenishment Process and Batched Replenishment Process under a Stochastic Replenishment Process}
When the replenishment process is stochastic, recall that each replenishment
$\zeta_i(j)$ is a non-negative random variable supported on $[0,M]$ with
expected value $q_{ij}$, where $M\geq1$. Our goal is to replace this
stochastic replenishment process with a fully deterministic batched
replenishment process without observing any realizations.

As described above, we first define the fluid replenishment process $\{\zeta_i^F(j)\}_{i \in \mathcal{I}, j \in \mathcal{J}}$ where each $\zeta_i^F(j)$ is set to $(1-\epsilon) q_{ij}$. Here, $\epsilon$ is defined as:
\begin{equation*}
    \epsilon = \min \left( \left(\frac{3M}{c_{\min}}\log(c_{\min} d)\right)^\frac{1}{3} , 1\right).
\end{equation*} 
Note that in the large inventory regime with
$M=\mathcal{O}(c_{\min}^{1-\nu})$ for some constant $\nu>0$, $\epsilon$ approaches $0$. As shown in Section \ref{Analysis}, this specific choice of $\epsilon$ guarantees that the batched replenishment process (which we will define shortly) can be implemented without any deficiency with high probability, by leveraging well-known concentration bounds. 

This fluid replenishment process $\{\zeta_i^F(j)\}_{i \in \mathcal{I}, j \in \mathcal{J}}$ can be computed before the arrival of the first request. Then, we define the corresponding batched replenishment process $\{\zeta_i^B(j)\}_{i \in \mathcal{I}, j \in \mathcal{J}}$ similarly to that in Section \ref{3.1.1.}, but with a different batching threshold: $c_{B} = \epsilon c_{\min}$.\footnote{Recall that the batching threshold is set to $\sqrt{c_{\min}}$ under an adversarial replenishment process.} This batched replenishment process $\{\zeta_i^B(j)\}_{i \in \mathcal{I}, j \in \mathcal{J}}$ is also computed in advance, with each replenishment being either 0 or at least $c_B$.

\subsubsection{Batched Instance and Batching Extension}

For each $i \in \mathcal{I}$ and $j \in \mathcal{J}$, if a batched replenishment, $\zeta_i^B(j)$ is positive (meaning $\zeta_i^B(j) \geq \epsilon c_{\min})$, we create a corresponding new resource type $i_j$ with starting inventory $\zeta_i^B(j)$. We also generate duplicated actions $k(i_j)$ for each action $k$, as defined in Section \ref{3.1.2.}. The resulting instance, which we call a batched instance, is denoted by $H_B \in \mathcal{H}_0$ and is detailed in Algorithm $\ref{alg:H_B_S}$.

\begin{algorithm}[h!]
\caption{Construction of the batched instance under stochastic processes}\label{alg:H_B_S}
\begin{algorithmic}
\State \textbf{Input:} An instance $H$ under stochastic arrival and replenishment processes
\State {Set $c_{\min} := \min_{i \in \mathcal{I}} c_i$,  $\epsilon := \min \left( \left(\frac{3M}{c_{\min}}\log(c_{\min} d)\right)^\frac{1}{3} , 1\right)$, and $c_B := \epsilon c_{\min}$.}
\State {Initialize $\mathcal{K}(0) := \mathcal{K}$.}

\For {each resource type $i \in \mathcal{I}$}
    \State {Initialize $\bar{c}_i := 0$.}
\EndFor

\For {each request $j \in \mathcal{J}$} \Comment{Update the batched replenishment process}
    \State {Initialize $\mathcal{K}_0(j) := \mathcal{K}(j-1)$.}
    \For {each resource type $i \in \mathcal{I}$}
        \State Set $\mathcal{K}_i(j) := \mathcal{K}_{i-1}(j)$.
        \State Set $\zeta_i^F(j) := (1-\epsilon) q_{ij}$. \Comment{Compute the fluid replenishment process first}
        \State $\bar{c}_i := \bar{c}_i + \zeta_i^F(j).$
        \If{ $\bar{c}_i \geq c_B$} \Comment{Create an identical resource type and actions}
        \State $\zeta_i^B(j) := \bar{c}_i$.
        \State $\mathcal{I} := \mathcal{I} \cup \{i_j\}$.
            \State $c_{i_j} := \bar{c}_i$.
        \State $\mathcal{K}_i(j) := \mathcal{K}_i(j) \cup \{k(i_j): k \in \mathcal{K}_i(j)\}$. 
            \State $\bar{c_i} := 0$. 
        \Else 
        \State {$\zeta_i^B(j) := 0.$}
        \EndIf
    \EndFor
    \State $\mathcal{K}(j) := \mathcal{K}_n(j)$. 
\EndFor
\State \textbf{return} Instance $H_B \in \mathcal{H}_0$ with updated $\mathcal{I}$ and $\mathcal{K}(m)$
\end{algorithmic}
\end{algorithm}
Note that our choice of $\epsilon$ and $c_B=\epsilon c_{\min}$ balances
the minimum initial inventory in the batched instance against the loss from
batching, while also accounting for the realization of the expected
replenishment with high probability. We formally analyze this trade-off in
Section~\ref{Analysis}.

The batching extension, ALG-B, runs ALG on the pre-computed $H_B$ instead of the original instance, $H$. Although $H_B$ no longer has exogenous replenishment, we must still account for the actual, random realizations of exogenous replenishment when implementing chosen actions. The main issue is that ALG, now running on $H_B$, makes decisions based on inventory derived from the batched replenishment process, whereas the actual replenishment may deviate from this projection, and it may lead to situations where the actual physical inventory is insufficient to execute an action planned by ALG on $H_B$. However, our analysis in Section \ref{Analysis} demonstrates that the probability of such an inventory shortfall is vanishingly small due to the $(1-\epsilon)$ term in our fluid replenishment process definition.

Despite this low probability, such capacity violations require a specific handling mechanism. When the amount of realized replenishment is insufficient to match the resource availability assumed by ALG's decision on $H_B$, we replace the intended action with a trivial action $k_0$ (meaning, no resources are actually allocated). Crucially, to maintain consistency with ALG's decision on $H_B$, we let ALG operate under the assumption that the original, intended action was successfully implemented. This means that while the resources are not consumed, ALG's ``virtual'' inventory and resource usage are updated as if the original action occurred. This creates a discrepancy between the ``virtual'' execution that ALG (on $H_B$) perceives and the ``real'' execution that ALG-B (on $H$) performs.  

As a result, ALG-B's virtual execution (what ALG imagines) allows ALG to operate as if it were simply running on $H_B$. Its real execution, however, incorporates a fallback to action $k_0$ only when capacity violations occur due to insufficient replenishment, matching the resource availability. This strategy enables us to leverage ALG's performance guarantees from the fixed-inventory setting, provided we can bound the probability that such a fallback happens\footnote{If $M$ is of $\Omega(c_{\min})$, we cannot bound the probability in general, corresponding to the impossibility result of Theorem \ref{thm4.3}.}. The resulting algorithm, ALG-B, for instances under stochastic arrival and replenishment processes, is described in Algorithm \ref{alg:ALG-B_S}. We refer readers to Appendix \ref{appendix-ex} for an illustrative example.



\begin{algorithm}[h!]
\caption{Definition of $\text{ALG-B}$ under stochastic processes}\label{alg:ALG-B_S}
\begin{algorithmic}
\State \textbf{Input:} An instance $H$ under stochastic arrival and replenishment processes
\State {Calculate $H_B$ based on Algorithm $\ref{alg:H_B_S}$.} \Comment{Pre-compute the batched instance}
\State {Initialize $K_{past} = [\quad]$.} \Comment{Sequence of actions chosen by ALG on $H_B$}
\For {each request $j \in \mathcal{J}$} \Comment{Process arrivals online}
\State Choose an action $k(j)$ using ALG on $H_B$, based on $K_{past}$ and the observed filtration.
\State Append $k(j)$ to $K_{past}$.
\If{$k(j)$ is implementable under the actual realization}
\State Implement the original action in $\mathcal{K}$ corresponding to $k(j)$. \Comment{Real execution}
\Else
\State Implement $k_0$. \Comment{Real execution failed}
\EndIf
\State Update the filtration as if $k(j)$ had been implemented. \Comment{Virtual execution}
\EndFor
\end{algorithmic}
\end{algorithm}

\section{Analysis of the Batching Extensions}\label{Analysis}

In this section, we present the proof establishing the near-equivalence of the competitive ratio of any given algorithm $\text{ALG}$ on instances with fixed capacity and that of its batching extension, $\text{ALG-B}$, on instances with exogenous replenishment. 

We start by proving Theorem $\ref{thm4.1}$, then extend its proof technique to Theorem $\ref{thm4.2}$, and finally prove the impossibility result corresponding to Theorem $\ref{thm4.3}$. Formally, we revisit Theorem~\ref{thm4.1} from the main results.

\noindent\textbf{Theorem~\ref{thm4.1}} \textit{Given a family of instances $\mathcal{H}$ with exogenous replenishment and an adversarial arrival process, suppose that ALG achieves a competitive ratio of at least $\alpha(c)$ for $\mathcal{H}_0(c)$. There exists an algorithm, ALG-B, that achieves a competitive ratio of at least $\alpha\left(\sqrt{c}\right)\left(1-\mathcal{O}\left(\sqrt{\frac{\log(cd)}{c}}\right)\right)$ for $\mathcal{H}(c)$.}

\subsection{Proof of Theorem \ref{thm4.1}}\label{sProofIdea}

Recall that the result with both adversarial arrival and adversarial replenishment processes naturally extends to instances with an adversarial arrival process and a stochastic replenishment process. This is because, in the latter case, we can simply disregard the prior information about the stochastic replenishment and treat it as adversarial. Therefore, to establish Theorem \ref{thm4.1}, our primary focus in the initial part of the analysis will be on proving the result for instances where both the arrival and replenishment processes are adversarial.

\begin{figure}[!h]
    \centering
    \includegraphics[width=0.999\linewidth]{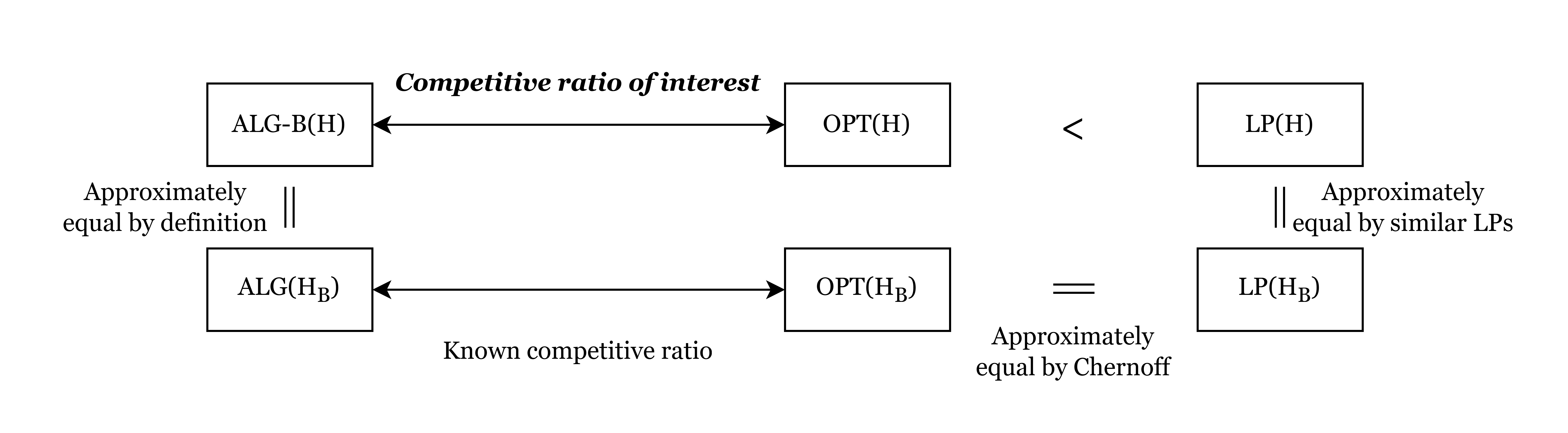}
    \caption{Analysis of $\text{ALG-B}$ with batching}
    \label{fig:batching}
\end{figure}

The high-level idea of the analysis is to bound the ratio of $\text{ALG-B} (H)$ to $\text{OPT} (H)$ using the known competitive ratio $\alpha(c_B)$ of $\text{ALG} \left(H_{B}\right)$ to $\text{OPT} \left(H_{B}\right)$. To analyze the impact of batching on the offline benchmark, we introduce an expected LP formulation, which provides a tractable and comparable relaxation for both instances $H$ and $H_B$, as intermediate benchmarks to bridge $\text{OPT} (H)$ and $\text{OPT} \left(H_B\right)$. The overall idea of analysis is illustrated in Figure \ref{fig:batching}.

Now, we are ready to prove Theorem \ref{thm4.1}. We first compare the benchmarks of the original instance $H$ and the transformed instance $H_B$. In particular, note that we can express the competitive ratio of the batching algorithms as follows:

\begin{align*}
    \min_{H \in \mathcal{H}(c)} \frac{\text{ALG-B}(H)}{OPT(H)} & = \min_{H \in \mathcal{H}(c)} \frac{\text{ALG-B}(H)}{\text{ALG}(H_B)} \frac{\text{OPT}(H_B)}{\text{OPT}(H)} \frac{{\text{ALG}(H_B)}}{\text{OPT}(H_B)}.
\end{align*}

By the definition of ALG-B, ALG-B$(H)$ = ALG$(H_B)$. Also, the minimum starting inventory of the transformed instance $H_B$ is at least $c_B$, which is $\sqrt{c}$ in this setting, because we can regard each batched exogenous replenishment as a starting inventory for a new resource type. Recall that we denote the competitive ratio of any algorithm $ALG$ by $\alpha(c)$ for the family of no-replenishment instances where the initial inventory is at least $c$. Therefore, we have that for any $H \in \mathcal{H} (c)$, the following inequality holds.

\begin{align*}
    \frac{\text{ALG-B}(H)}{OPT(H)} & \ge \frac{\text{ALG-B}(H)}{\text{ALG}(H_B)} \frac{\text{OPT}(H_B)}{\text{OPT}(H)} \min_{H_B \in \mathcal{H}_0 \left(\sqrt{c}\right)} \frac{{\text{ALG}(H_B)}}{\text{OPT}(H_B)} \geq \alpha \left(\sqrt{c}\right) \frac{\text{OPT}(H_B)}{\text{OPT}(H)}.
\end{align*}

Then we establish a formal definition for the expected LP whose optimal objective is approximately equal to the offline optimal. We unify the expected LP formulations for instances under both adversarial and stochastic settings. Note that when the replenishment process is adversarial, without loss of generality, request $j$ is assumed to be of type $j$, i.e. $p_{jj} = 1$, for every $j \in \mathcal{J}$ and denote realized replenishment by $\zeta_i(j) = q_{ij}$. 

Let \(a_{i j k z}(l)\) denote the expectation of \(A_{i j k z}(l)\) and \(r_{i j k z}\) denote the expectation value of \(R_{i j k z}\), respectively. Given an instance \(H\), the following formulation (LP$(H)$) can be used to get an upper bound of OPT$(H)$:

%
%

\begin{align}
    \left(\text{LP}\left(H\right)\right)  \quad \max \quad & \lambda  \quad &  \label{o1LP}\\
    \quad \text{s.t.} \quad & \lambda \leq \sum\limits_{j \in \mathcal{J}}\sum\limits_{k \in \mathcal{K}}\sum\limits_{z \in Z} r_{ijkz} x_{jkz}, \quad \forall i \in \mathcal{I}, \label{c1LP}\\
    & \sum\limits_{j \in [l]}\sum\limits_{k \in \mathcal{K}}\sum\limits_{z \in Z} a_{i j k z}(l) x_{ jkz} \leq c_i + \sum\limits_{j \in [l]} q_{ij}, \quad \forall i \in \mathcal{I}, l \in \mathcal{J},  & \label{c2LP}\\
    & \sum\limits_{k \in \mathcal{K}} x_{jkz} \leq p_{jz}, \quad \forall j \in \mathcal{J}, z \in Z \label{c3LP}\\
    & x_{jkz} \geq 0, \quad \forall j \in \mathcal{J}, k \in \mathcal{K}, z \in Z. \label{c4LP}
\end{align}

In the original problem, the variable $x_{jkz}$ can be interpreted as a binary decision variable that takes a value $1$ only if action $k$ is chosen for the request $j$ of request type $z$, and only one action can be chosen for each request. In (LP($H$)), however, this condition is relaxed as shown in (\ref{c3LP}) and (\ref{c4LP}). Additionally, the usage of resources and the reward for each action are deterministic. The objective (\ref{o1LP}) and constraint (\ref{c1LP}) define $\lambda$ as the minimum expected reward among all resource types, and constraint (\ref{c2LP}) ensures that the total amount of resource type $i$ used does not exceed the cumulative inventory for any resource type.

Let LP$(H)$ denote the optimal objective function value of (LP$(H)$). Recall that OPT$(H)$ represents the expected reward achieved by the optimal policy for instance \(H\). The optimal policy can be represented as \((X_{jkz}(\omega_j))_{j \in \mathcal{J}, k \in \mathcal{K}, z \in Z }\), where \(X_{jkz}\) has a value of 1 if action \(k\) is selected for request $j$ of type $z$ and has a value of 0 otherwise, based on the filtration $\omega_j$ up to the arrival of request $j$. Consequently, OPT$(H)$ can be expressed as \(\min\limits_{i \in \mathcal{I}} \left(\sum\limits_{j \in \mathcal{J}} \sum\limits_{k \in \mathcal{K}} \sum\limits_{z \in Z} r_{ijkz } \mathbb{E}[X_{jkz}(\omega_j)]\right)\). The following lemma establishes that the linear program (LP) provides an upper bound to the expected optimal reward.

\begin{lemma}\label{L2.1}
    For any instance $H$, $\text{LP}(H) \geq \text{OPT}(H)$.
\end{lemma}

Next, we show that as the initial inventory \(c_i\) increases, the optimal reward $\text{OPT}(H)$ approaches $\text{LP}(H)$. This result leverages the attenuated independent rounding technique and the well-known concentration bounds. Recall that \(c_{\min} = \min\limits_{i \in \mathcal{I}} c_i \) is the minimum initial inventory across all resource types and $d$ is the maximum number of resource types affected by any action $k \in \mathcal{K}$. Lemma \ref{L2.2} shows that the gap between $\text{LP}(H)$ and $\text{OPT}(H)$ depends on $c_{\min}$ and $d$. The full proof of Lemma \ref{L2.1} and Lemma \ref{L2.2} is given in Appendix \ref{appendix-proof-lp}.

\begin{lemma}\label{L2.2}
    For any instance $H$ with a deterministic replenishment process, \[\text{OPT}(H) \geq \left(1- \mathcal{O}\left(\sqrt{\frac{\log (c_{\min}d)}{c_{\min}}}\right)\right) \text{LP}(H).\]
\end{lemma}

Lastly, consider the expected LP of the instance $H_B$, which can be written as:

\begin{align*}
    \left(\text{LP}\left(H_B\right)\right) \quad \max \quad & \lambda  \quad &\\
    \quad \text{s.t.} \quad & \lambda \leq \sum\limits_{j \in \mathcal{J}}\sum\limits_{k \in \mathcal{K}} \sum_{z \in Z} r_{ijkz} x_{jkz}, \quad \forall i \in \mathcal{I},\\
    & \sum\limits_{j \in [l]}\sum\limits_{k \in \mathcal{K}} \sum_{z \in Z}  a_{i j k z}(l) x_{j k z} \leq c_i + \sum\limits_{j \in [l]} \zeta_i^B(j), \quad \forall i \in \mathcal{I}, l \in \mathcal{J},\\
    & \sum\limits_{k \in \mathcal{K}} x_{jkz} \leq p_{jz}, \quad \forall j \in \mathcal{J}, z \in Z\\
    & x_{jkz} \geq 0, \quad \forall j \in \mathcal{J}, k \in \mathcal{K}, z \in Z.
\end{align*}

By the definition of the batching extension, we have the following inequality:

\begin{equation*}
    \sum\limits_{j \in [l]} q_{ij} - \sqrt{c} \leq \sum\limits_{j \in [l]} \zeta_i^B(j) \leq \sum\limits_{j \in [l]} q_{ij}, \quad \forall i \in \mathcal{I}, l \in \mathcal{J},
\end{equation*}
because the batching strategy ensures that the cumulative unbatched replenishment never exceeds $c_B$.

Therefore, LP$(H_B)$ is bounded below by $\left(1-\frac{1}{\sqrt{c}}\right) \text{LP}(H)$. By Lemma \ref{L2.1} and \ref{L2.2},
\begin{align*}
\frac{\text{ALG-B} \left(H\right)}{{\text{OPT}} \left(H\right)} & \ge \alpha \left(\sqrt{c}\right) \left(1 - \mathcal{O}\left(\sqrt{\frac{\log\left( cd \right)}{c}}\right)\right) \frac{{\text{LP}} \left(H_B\right)}{{\text{LP}} \left(H\right)}. \\
& = \alpha \left(\sqrt{c}\right) \left(1 - \mathcal{O}\left(\sqrt{\frac{\log\left( cd \right)}{c}}\right)\right).
\end{align*}
\hfill{$\square$}


\subsection{Proof of Theorem \ref{thm4.2}}

In this section, we extend the proof techniques used in Theorem \ref{thm4.1}—which addresses the adversarial arrival setting—to accommodate stochastic arrival and replenishment processes, as considered in Theorem \ref{thm4.2}. Although the specifics differ due to the stochastic nature of these processes, the core principles behind the batching strategy and its analysis remain effective and are adapted accordingly.

\noindent\textbf{Theorem 2.2} \textit{Given a family of instances $\mathcal{H}$ and $c>0$ under a stochastic arrival process and a stochastic replenishment process, suppose that $\text{ALG}$ achieves a competitive ratio of at least $\alpha(c)$ for the set of instances $\mathcal{H}_{0}(c)$. 
There exists an algorithm, ALG-B, that achieves a competitive ratio of at least $\alpha\left(
\mathcal{O}\left(
M^{\frac{1}{3}} c^{\frac{2}{3}} \log^{\frac{1}{3}}(cd)
\right)\right)
\left(
1-\mathcal{O}\left(
M^{\frac{1}{3}} c^{-\frac{1}{3}} \log^{\frac{1}{3}}(cd)
\right)
\right)$ for the set of instances $\mathcal{H}(c)$. }

Recall the analysis in Section 4.1 and the proof sketch in Figure \ref{fig:batching}. We first derive that ALG-B$(H)$ is approximately equal to ALG$(H_B)$. Unlike in the adversarial setting, where the two are equivalent, a gap arises when the realizations of exogenous replenishment are insufficient to implement an action chosen by ALG on pre-computed batched instance $H_B$.

Consider a request $j$, and suppose that ALG selects action $k \in \mathcal{K}$ for this request in the deterministic instance $H_B$. The batching extension, ALG-B, defined in Algorithm \ref{alg:ALG-B_S}, attempts to take the same action $k$ for request $j$, provided that in the actual instance $H$ the cumulative realized exogenous replenishment for the relevant resources is sufficient to match the amount that action $k$ requires in $H_B$. Let $\mathcal{I}_k \subset \mathcal{I}$ be the set of resource types associated with action $k$ with $|\mathcal{I}_k| \leq d$. Fix a resource type for accounting rewards and denote expected rewards (associated with the resource type) obtained from request $j$ by ALG-B$(H)$ and ALG$(H_B)$ as $\text{ALG-B}_j(H)$ and $\text{ALG}_j(H_B)$, respectively. By linearity, we have that
\begin{equation*}
    \text{ALG-B}(H) = \sum\limits_{j \in \mathcal{J}} \text{ALG-B}_j(H), \quad \text{ALG}(H_B) = \sum\limits_{j \in \mathcal{J}} \text{ALG}_j(H_B), 
\end{equation*}
and
\begin{align*}
    \text{ALG-B}_j(H) & \geq \mathbb{P} \left(\sum\limits_{t \in [j]} \zeta_i(t) \geq \sum\limits_{t \in [j]} \zeta_i^B(t) \text{for all } i \in \mathcal{I}_k\right) \text{ALG}_j(H_B) \\
    & \geq \left(1 - \sum\limits_{i \in \mathcal{I}_k} \mathbb{P} \left( \sum\limits_{t \in [j]} \zeta_i(t) < \sum\limits_{t \in [j]} \zeta_i^B(t) \right) \right) \text{ALG}_j(H_B).
\end{align*}

Note that for any $i \in \mathcal{I}$ and $j \in \mathcal{J}$, by the definition of the batching extension and the Chernoff bound (see Lemma \ref{Chernoff} in Appendix \ref{appendix-proof-lp} for the details), 
\begin{align*}
    \mathbb{P} \left( \sum\limits_{t \in [j]} \zeta_i(t) < \sum\limits_{t \in [j]} \zeta_i^B(t) \right) & \leq \mathbb{P} \left( \sum\limits_{t \in [j]} \zeta_i(t) < \sum\limits_{t \in [j]} (1- \epsilon) q_{it}, \sum\limits_{t \in [j]} (1 - \epsilon) q_{it} \geq \epsilon c\right) \\
    & \leq \exp\left(-\frac{\epsilon^3c}{3M}\right) = \frac{1}{cd},
\end{align*}
where $\epsilon = \left(\frac{3M}{c} \log (cd) \right)^\frac{1}{3} < 1$ in the asymptotic regime.

Therefore,
\begin{equation*}
\text{ALG-B}_j(H) \geq \left(1-\frac{1}{c}\right) \text{ALG}_j(H_B), \quad \forall j \in \mathcal{J},
\end{equation*}
and since this inequality holds for every resource type for accounting rewards, taking the minimum over all resource types yields
\begin{equation*}
\text{ALG-B}(H) \geq \left(1-\frac{1}{c}\right) \text{ALG}(H_B). 
\end{equation*}

Similar to the analysis in Section 4.1, the next step is to show that $\text{OPT} \left(H_B\right)$ is approximately equal to $\text{OPT} (H)$ by intermediate LP benchmarks. To compare (LP $(H)$) to (LP $(H_B)$), according to the definition of the batching extension, we have the following inequality:

\begin{equation*}
    \sum\limits_{j \in [l]} (1-\epsilon)q_{ij} - \epsilon c \leq \sum\limits_{j \in [l]} \zeta_i^B(j) \leq \sum\limits_{j \in [l]} (1-\epsilon) q_{ij}, \quad \forall i \in \mathcal{I}, l \in \mathcal{J},
\end{equation*}
implying $\text{LP} (H_B) \geq (1-\epsilon) \text{LP} (H)$. Finally, since $\frac{{\text{ALG}(H_B)}}{\text{OPT}(H_B)}$ is at least $\alpha(c_B) = \alpha(\epsilon c)$, using the result of Lemma \ref{L2.1} and \ref{L2.2},\footnote{Recall that Lemma~\ref{L2.2} applies to instances with deterministic replenishment. Although now $H$ has stochastic replenishment, $H_B$ has deterministic replenishment, so we can still apply Lemma~\ref{L2.2} to $H_B$.} we have
\begin{align*}
\frac{\text{ALG-B}(H)}{\text{OPT}(H)}
& =
\frac{\text{ALG-B}(H)}{\text{ALG}(H_B)}
\frac{\text{OPT}(H_B)}{\text{OPT}(H)}
\frac{\text{ALG}(H_B)}{\text{OPT}(H_B)} \\
& \geq
\left(1-\frac{1}{c}\right)
\left(
1-\mathcal{O}\left(
\sqrt{\frac{\log(c_Bd)}{c_B}}
\right)
\right)
(1-\epsilon)\alpha(\epsilon c).
\end{align*}

Since $c_B = \epsilon c$,
\[
\sqrt{\frac{\log(c_Bd)}{c_B}}
\leq
\sqrt{\frac{\log(cd)}{\epsilon c}}
=
\frac{\epsilon}{\sqrt{3M}}
=
\mathcal{O}(\epsilon). 
\]
Note that our choice of $\epsilon$ is designed so that the error term above
is also of order $\mathcal{O}(\epsilon)$, consistent with the
$(1-\epsilon)$ loss from batching.
Therefore,
\[
\frac{\text{ALG-B}(H)}{\text{OPT}(H)}
\geq
\alpha(\epsilon c)
\left(1-\mathcal{O}(\epsilon)\right).
\]
\hfill $\square$

\subsection{Proof of Theorem \ref{thm4.3}}

\noindent\textbf{Theorem 2.3} \textit{Fix any constant $\gamma\in(0,1]$. There exists a family of instances
$\mathcal{H}$ under a stochastic arrival process and a stochastic
replenishment process with $M=\gamma c$ such that the corresponding
setting without exogenous replenishment admits an online algorithm with
competitive ratio $1$, whereas every online algorithm for $\mathcal{H}$
has competitive ratio at most
\[
1-\frac{\gamma}{2+3\gamma}.
\]}

Fix $\gamma\in(0,1]$ and for simplicity, assume that $\gamma c$ is an
integer. We construct an instance with $M=\gamma c$. 
Consider the following base instance structure under the stochastic arrival setting, denoted by $G$: there is only one type of resource ($\mathcal{I} = \{1\}$) with initial inventory $c$, a total of $m = (1 + \gamma) c$ requests, and two request types ($Z = \{1, 2\}$). The requests arrive in sequence, indexed by $\mathcal{J} = [(1 + \gamma) c]$. Stochastic arrival request $j$ is realized as type $z$ with probability $p_{jz}$: for requests $j \in \{1, \ldots, c\}$, the type is always $1$ ($p_{j1} = 1, p_{j2} = 0$), and for requests $j \in \{c+1, \ldots, (1 + \gamma) c\}$, the type is always $2$ ($p_{j1} = 0, p_{j2} = 1$).

There is only one nontrivial action, $k$, which consumes a unit amount of resource $1$ with probability $1$. The set of actions is thus $\mathcal{K} = \{k, k_0\}$, where $k_0$ is the trivial action. If the nontrivial action $k$ is chosen for a request of type $1$, the reward is $1$. If $k$ is chosen for a request of type $2$, the reward is $2$. 

In the absence of replenishment, as the arrival sequence is known in
advance, an optimal policy is to reserve $\gamma c$
units of inventory for the last $\gamma c$ type-2 requests and to serve
$(1-\gamma)c$ of the preceding type-1 requests. Therefore, without replenishment, the instance admits a simple online algorithm with competitive ratio $1$.

Now, to prove Theorem \ref{thm4.3}, we introduce an instance based on $G$ with a stochastic exogenous replenishment process. Consider instance $G_S$, which uses the request and action structure of $G$ with initial inventory $c$. The replenishment process for $G_S$ is defined as follows: $\zeta_1(j) = 0$ for all $j \in \mathcal{J} \setminus \{c+1\}$. At time $c+1$, there is a single possible replenishment amount of $\gamma c$ units, but whether it arrives is uncertain to the online algorithm at earlier times. Specifically, $\zeta_1(c+1) = \gamma c$ with probability 0.5, and $\zeta_1(c+1) = 0$ with probability 0.5.

The high-level idea is to design a replenishment process in which the maximum possible replenishment amount of a single resource in a single replenishment, denoted by \( M \), is not small enough compared to \( c \), so that the resource accumulation required for effective batching is no longer guaranteed with high probability. In the instance \( G_S \), we construct \( M = \gamma c \) so that Theorem~\ref{thm4.2} does not guarantee the existence of a near-optimal batching extension.

\noindent\textbf{Offline Optimal Solution:} The offline optimal algorithm knows the realized replenishment outcome at time $c+1$ before making any allocation decisions. If no replenishment occurs (with probability 0.5), the optimal strategy is to allocate $\gamma c$ units to type-2 requests to earn reward 2 per request and remaining $(1 - \gamma) c$ units to type-1 requests to earn reward 1 per request. The total reward is $(1 + \gamma)c$. If replenishment of $\gamma c$ units occurs (with probability 0.5), all the $(1 + \gamma) c$ request can be served so the total reward is $(1 + 2\gamma) c$. Thus, the expected reward of the offline optimal algorithm for instance $G_S$ is $\left(1 + \frac{3}{2} \gamma\right) c$

\noindent\textbf{Best Online Algorithm:} Consider any online algorithm operating on instance $G_S$. The algorithm must decide on an allocation strategy for type-1 requests (arriving first) without knowing the realization of the replenishment at time $c+1$. Suppose an online algorithm adopts a strategy that, in expectation or deterministically, allocates $x$ units of the initial resource to serve type-1 requests (where $0 \le x \le c$). This yields a reward of $x$. The remaining initial inventory is $c-x$.

If no replenishment occurs at time $c+1$ (with probability $0.5$), only the remaining $c-x$ initial units are available for the $\gamma c$ type-2 requests, so the reward from type-2 requests is at most $2(c-x)$. The total reward in this case is at most $x + 2(c-x)$.

If replenishment of $\gamma c$ units occurs at time $c+1$ (with probability 0.5), the total inventory available for type-2 requests is $(1 + \gamma) c-x$. The algorithm can utilize up to $\gamma c$ units from the replenishment for all $\gamma c$ type-2 requests, gaining a reward of $2\gamma c$. The total reward in this case is at most $x + 2\gamma c$.

Therefore, the expected reward for any online algorithm is upper bounded by:
\[
\frac{1}{2} \left( x + 2(c - x) \right) + \frac{1}{2} \left( x + 2\gamma c \right) = \frac{1}{2} (x + 2c - 2x + x + 2\gamma c) = (1 + \gamma)c.
\]
Since this expected reward is independent of the specific value of $x$ chosen by the online algorithm (as long as $0 \le x \le c$), the maximum expected reward any online algorithm can achieve on instance $G_S$ is $(1+\gamma)c$. Thus, the competitive ratio of any online algorithm for instance $G_S$ is at most \[\frac{2 + 2\gamma}{2 + 3\gamma}<1.\]

\hfill $\square$

Based on the hard instance $G_S$ with stochastic arrivals and stochastic replenishment of a large amount of resources, we can construct a hard instance under stochastic arrivals and adversarial replenishment. Consider the following two instances, $G_1$ and $G_2$, based on $G$, which feature an adversarial replenishment process. The first instance, $G_1$, is the same as $G$ with the replenishment process being adversarial such that $\zeta_1(j) = 0$ for all $j \in \mathcal{J}$, i.e., there is no exogenous replenishment in this case.

The second instance, $G_2$, is also the same as $G$ with the replenishment process being adversarial such that $\zeta_1(j) = 0$ for all $j \in \mathcal{J} \setminus \{c+1\}$, and $\zeta_1(c+1) = \gamma c$. Then, by the above argument, for any online algorithm, the competitive ratio for at least one of instances $G_1$ or $G_2$ is at most $\frac{2 + 2\gamma}{2 + 3\gamma}$, asymptotically less than $1$ when $\gamma = \Omega (1)$.

\section{Leveraging Problem Structure in Batching Extensions}\label{SecStructure}

The results in the previous sections establish performance
guarantees for batching extensions across a broad class of online
resource allocation problems. It is natural to ask whether
the batching extension and its analysis can be further strengthened
in more structured settings. In particular, we consider two questions:
(i) Can we obtain a sharper convergence bound to the optimal competitive ratio by
exploiting additional problem structure, and if so, which step
of our general analysis can be improved?
(ii) In practice, decision-makers may prefer to utilize unbatched
inventory immediately rather than wait for a batch to form.
Can we accommodate such myopic allocations in certain structured
settings while preserving the competitive-ratio guarantee of ALG-B? 

In this section, we explore these questions and provide positive answers in online assortment optimization. 
It is worth noting that such refinements do not readily extend to the online resource allocation framework in Section~\ref{Notation}. In particular, the improved convergence relies heavily on problem-specific structure, and myopic allocations are suboptimal in general even without replenishment.\footnote{For example, even in classic single-resource revenue management settings \citep{talluri1998analysis,ball2009toward}, when rewards (fares) are nonstationary (e.g., arrival-dependent), optimal policies are necessarily non-myopic because capacity must be protected for higher-paying future arrivals. Moreover, when actions consume bundles of multiple resources, myopically depleting one resource can make profitable future bundles infeasible.}

We first show in Section~\ref{sec:assortment} that the dependence on the number of products can be removed from the performance guarantee by evaluating inventory availability for each product separately. Section~\ref{SecMyopic} exploits stationary rewards (each unit of product $i$ consumed earns the same reward $r_i$ across arrivals) to utilize unbatched inventory myopically while preserving the performance of ALG-B. More generally, the first idea applies whenever any action can be replaced by a (randomized) action on the available resources that preserves each available resource's marginal consumption probability, and the expected reward decomposes additively across resources. The second relies on stationary resource-specific rewards and also extends to reusable resources under stationary usage distributions.

Formally, we consider the setting of \citet{golrezaei2014real} under adversarial
arrival and replenishment processes. There are $n$ products, indexed by
$\mathcal I=[n]$, with initial inventories $c_i \geq 0$ and rewards $r_i\geq0$
per unit sold. Also, recall that we define $c_{\min} := \min_{i \in \mathcal{I}} c_i$.
Under the adversarial replenishment process, with customer $j\in\mathcal J$, $\zeta_i(j) \geq 0$ new units of product
$i$ arrive. Without loss of generality, customer $j$ is of type $j$.
An action is an assortment $S\in\mathcal S=2^{\mathcal I}$, with
$k_0=\emptyset$. Customer $j$ selects at most one product, choosing
$i\in S$ with probability $\phi^j(i,S)$ and the no-purchase option
$0$ with probability $\phi^j(0,S)=1-\sum_{i\in S}\phi^j(i,S)$.
Conditional on the offered assortment, each customer's choice is
independent of the past. We set $\phi^j(i,S)=0$ for $i\notin S$ and assume the
standard substitutability property:
\[
\phi^j(i,S')\geq\phi^j(i,S),
\qquad S'\subseteq S,\quad i\in S'.
\]

For simplicity, we allow unavailable products to be shown to customers:
the customer's choice is observed, and if the selected product $i$ is unavailable,
the customer leaves without purchasing; otherwise, one unit of product $i$
is consumed and revenue $r_i$ is earned.
This assumption is without loss of generality even when only available products can be displayed, as a randomized assortment
of available products can preserve the original marginal purchase probabilities.\footnote{
Under the substitutability assumption, we can replace any assortment
$S$ by a randomized assortment $\widetilde S\subseteq S\cap U$,
where $U$ is the set of currently available products, satisfying
\[
\mathbb E\bigl[\phi^j(i,\widetilde S)\bigr]=\phi^j(i,S),
\qquad i\in S\cap U,
\]
as shown by \citet{goyal2025asymptotically} and
\citet{feng2024near}.
Incorporating this construction as a subroutine preserves the marginal
purchase probabilities of all available products and hence yields
the same expected revenue and performance guarantee.
}
The objective is to maximize expected total revenue.

Without replenishment, the Inventory-Balancing algorithm
\citep{golrezaei2014real} achieves
$\alpha(c)=(1-1/e)(1-\mathcal O(1/c))$ for instances with $c_{\min} \geq c$.
Let ALG denote this algorithm and ALG-B its batching extension.
Since an assortment may contain $n$ products, directly applying
Theorem~\ref{thm4.1} with $d=n$ gives
\[
\left(1-\frac1e\right)
\left(1-\mathcal O\left(\sqrt{\frac{\log(nc)}{c}}\right)\right).
\]

Note that ALG-B attains the optimal asymptotic competitive ratio $1-1/e$, but the performance guarantee depends on $n$, which does not appear in the previous work of \cite{feng2021robustness}. Also, ALG-B always ignores the unbatched replenishment, missing a potential opportunity and may not be preferable in the real-world application.

\subsection{Strengthened Approximation Ratio of the LP Relaxation}\label{sec:assortment}

In our general black-box analysis, the dependence on $n$ arises from the union bound in Lemma~\ref{L2.2}, since an action can be implemented only if all of its required resources are simultaneously available. In online assortment optimization, however, an assortment $S$ can be offered even when some products in $S$ are out of stock, and the expected reward decomposes additively across products. Consequently, earning revenue from product $i$ requires only that product $i$ itself be available when chosen, regardless of the availability of other products in $S$. Bounding the stockout probability for each product separately thus avoids the union bound over the assortment and yields the following strengthening of Lemma~\ref{L2.2}.

\begin{lemma}\label{lem:assortment-lp}
For any instance $H$ of the online assortment problem with a
deterministic replenishment process,
\[
\mathrm{OPT}(H)\geq
\left(1-\mathcal O\left(
\sqrt{\frac{\log c_{\min}}{c_{\min}}}\right)\right)
\mathrm{LP}(H).
\]
\end{lemma}

\begin{proof}
The proof is analogous to the original proof of Lemma~\ref{L2.2}. Let $(x^*_{jS})$ be an optimal solution of $\mathrm{LP}(H)$.
Independently for each customer $j$, sample assortment $S$ with
probability $x^*_{jS}/(1+\delta)$, where $\delta \in (0, 1]$ is a parameter to be chosen later, and offer $\emptyset$ with the
remaining probability. Note that we provide the assortment even if some products are unavailable. Let $X_{ij}$ indicate that the
potential choice of customer $j$ is product $i$. For each fixed $i$, these indicators
are independent across $j$, with
\[
\mathbb E[X_{ij}]
=\frac{1}{1+\delta}\sum_{S\in\mathcal S}x^*_{jS}\phi^j(i,S).
\]

Let $U_j$ denote the available products before customer $j$
chooses. Since actual sales never exceed potential choices,
$i\notin U_j$ implies $\sum_{t<j}X_{it}\geq c_i+\sum_{t\leq j}\zeta_i(t)$. Therefore, for $\delta=\sqrt{6\log(c_{\min})/c_{\min}}$,
the Chernoff bound yields
\[
\mathbb P(i\notin U_j)
\leq\exp\left(-\frac{\delta^2 c_{\min}}6\right)
\leq\frac1{c_{\min}}.
\]
Since $X_{ij}$ is independent of $U_j$, expected total revenue is at least
\[
\sum_{j\in\mathcal J}\sum_{i\in\mathcal I}
r_i\mathbb E[X_{ij}]\mathbb P(i\in U_j)
\geq\frac{1-1/c_{\min}}{1+\delta}\,\mathrm{LP}(H),
\]
which proves the claim for sufficiently large $c_{\min}$.
\end{proof}

Finally, as in the proof of Theorem~\ref{thm4.1} with Lemma~\ref{lem:assortment-lp}, we obtain
\[
\frac{\mathrm{ALG\mbox{-}B}(H)}{\mathrm{OPT}(H)}
\geq\alpha(\sqrt c)
\left(1-\mathcal O\left(\sqrt{\frac{\log c}{c}}\right)\right)
=\left(1-\frac1e\right)
\left(1-\mathcal O\left(\sqrt{\frac{\log c}{c}}\right)\right).
\]
This removes the dependence on $n$, consistent with prior results
\citep{golrezaei2014real,feng2021robustness}.

\subsection{Myopic Use of Unbatched Inventory}\label{SecMyopic}

Recall that ALG-B withholds replenished inventory until it forms a batch of size
at least $c_B=\sqrt{c_{\min}}$. Consequently, ALG-B may not offer any products (or may only offer unavailable products) even when unbatched inventory is available.
We construct a myopic extension, ALG-M, that uses this inventory
without reducing the performance of ALG-B.

The main idea behind ALG-M is simple: the algorithm follows the decisions of ALG-B whenever possible, but if ALG-B withholds unbatched inventory and fails to offer any available products, ALG-M departs from ALG-B and myopically offers an assortment of available products while keeping ALG-B oblivious to this departure.

To implement this idea, ALG-M maintains a virtual execution of ALG-B, while adding all
replenishment to its real inventory immediately. Let $S^{\mathrm B}(j)$
be the virtual assortment selected by ALG-B for customer $j$, and let $U_j^{\mathrm M}$ be the products available
in the real inventory. Normally, ALG-M offers $S^{\mathrm B}(j)$ and updates both executions with the customer's realized choice. However, if
$S^{\mathrm B}(j)\cap U_j^{\mathrm M}=\emptyset$
(including when $S^{\mathrm B}(j)=\emptyset$)
and an unbatched unit is available, ALG-M instead offers
a fallback assortment of available products containing a product
with an available unbatched unit, while updating the virtual execution of ALG-B by independently simulating the customer's choice under $S^{\mathrm B}(j)$. This fallback assortment may be chosen greedily; its particular choice does not affect the dominance below. The complete procedure is given in Algorithm~\ref{alg:ALG-M}.

\begin{algorithm}[H]
\caption{Definition of the myopic batching extension ALG-M}\label{alg:ALG-M}
\begin{algorithmic}
\For{each customer $j\in\mathcal J$}
    \State Add all new replenishment to the real inventory.
    \State Update the virtual execution of ALG-B and obtain $S^{\mathrm B}(j)$.
    \State Set $S^{\mathrm M}(j):=S^{\mathrm B}(j)$.
    \If{$S^{\mathrm B}(j)\cap U_j^{\mathrm M}=\emptyset$ and an unbatched unit is available}
        \State Replace $S^{\mathrm M}(j)$ by any assortment of available products containing a product with an available unbatched unit.
    \EndIf
    \State Offer $S^{\mathrm M}(j)$ and update real inventory after the customer's choice.
    \State Update virtual ALG-B using the same choice if $S^{\mathrm M}(j)=S^{\mathrm B}(j)$; otherwise, independently simulate the choice under $S^{\mathrm B}(j)$.
\EndFor
\end{algorithmic}
\end{algorithm}

The following lemma shows that utilizing unbatched inventory in a myopic way never reduces the expected total reward relative to $\text{ALG-B}$. Consequently, replacing $\text{ALG-B}$ with $\text{ALG-M}$ yields the same guarantee of at least $\alpha(\sqrt{c})\bigl(1-\mathcal{O}(\sqrt{\log c / c})\bigr)$.

\begin{lemma}\label{lem:myopic}
For any instance $H$ of the online assortment problem and any base
algorithm ALG, the myopic batching extension satisfies
\[
\mathrm{ALG\mbox{-}M}(H)\geq\mathrm{ALG\mbox{-}B}(H).
\]
\end{lemma}

\begin{proof}
By construction, we couple customer choices identically whenever the two executions offer
the same assortment, and simulate the virtual choice independently
otherwise. 
Let $p_i^{\mathrm M}(j)$ and $p_i^{\mathrm B}(j)$ denote the cumulative sales of product $i$ through customer $j$ in ALG-M and the simulated ALG-B, respectively. Since $\mathrm{ALG\mbox{-}M}(H) = \mathbb E[\sum_{i\in\mathcal I} r_i p_i^{\mathrm M}(m)]$ and $\mathrm{ALG\mbox{-}B}(H) = \mathbb E[\sum_{i\in\mathcal I} r_i p_i^{\mathrm B}(m)]$ with $r_i \geq 0$, it is sufficient to show that $p_i^{\mathrm B}(j) \leq p_i^{\mathrm M}(j)$ for all $i \in \mathcal I$ and $j \in \mathcal J$ on every coupled sample path.

Suppose, for contradiction, that $j$ is the first
customer at which $p_i^{\mathrm M}(j)<p_i^{\mathrm B}(j)$ for some $i$. Immediately before $j$, the two sales counts for $i$ must be equal.
Since cumulative real replenishment is at least cumulative batched replenishment,
if product $i$ is available virtually, it is then also available in the
real inventory. Thus, if ALG-B sells $i$ at $j$, then
$S^{\mathrm B}(j)\cap U_j^{\mathrm M}\ne\emptyset$: ALG-M offers the
same assortment and sells the same product under the coupled choice,
a contradiction. Hence $p_i^{\mathrm M}(j)\geq p_i^{\mathrm B}(j)$ for
all $i,j$, and
\[
\mathrm{ALG\mbox{-}M}(H)
=\mathbb E\left[\sum_i r_i p_i^{\mathrm M}(m)\right]
\geq\mathbb E\left[\sum_i r_i p_i^{\mathrm B}(m)\right]
=\mathrm{ALG\mbox{-}B}(H).
\]
\end{proof}

\textbf{Remark. } The same dominance extends to reusable products with independent, stationary, product-specific usage durations, using the FIFO queue
coupling of \citet{gong2022online}, i.e., assigning the same usage duration to the $q$th sale of each product in both executions for each $q \in \mathbb{N}$.

\section{Conclusion and Open Problems}\label{Conclusion}

We consider a general framework for incorporating exogenous replenishment in online resource allocation. Our black-box approach leverages existing algorithms and known performance guarantees for settings without replenishment to obtain algorithms with the same guarantees for settings with replenishment. The key idea is to batch new inventory of a resource until it reaches a threshold value, at which point we plug in all the new inventory as a ``new'' resource in the online algorithm. 

We analyze the competitive ratio of our approach in two steps: $(i)$ we compare the algorithm with batching against an offline benchmark that receives replenishment in batches and $(ii)$ we compare the benchmark with an LP relaxation of the offline problem. Step $(i)$ reduces to the competitive ratio of the original algorithm (without replenishment). When the starting inventory is sufficiently large, we show that the intermediate benchmark and LP benchmark have the same optimal value (Step $(ii)$). 

An interesting open question for future work is to extend our framework beyond non-negative, resource-specific replenishment. For example, negative inventory shocks, such as inventory loss due to perishability, prevent our monotonicity arguments and one-sided concentration bounds from applying directly. Likewise, allowing flexible replenishment that can be allocated across multiple compatible resource types may significantly alter the action set of the transformed instance, requiring new techniques beyond our current black-box reduction. Another challenging open problem is to remove the large inventory assumption and generalize (randomized) algorithms for the general inventory to settings with exogenous replenishment. Furthermore, in the setting of stochastic arrivals, while we have demonstrated the non-existence of a near-optimal online algorithm under stochastic replenishment with large replenishment quantities, the possibility of establishing a constant-factor performance guarantee remains an open question.

\bibliography{bibliographyJul19}


\newpage
\appendix
\section{Appendix}

\subsection{Illustrative Example of Batching Extension}\label{appendix-ex}

\paragraph{Example 1 (Under Adversarial Arrival and Replenishment Processes)} To illustrate how a batching extension dynamically constructs transformed instances, recall the instance with two types of resources \(\mathcal{I} = \{A, B\}\) with starting inventory \(c_A = c_B = 100\). Both the arrival and replenishment processes are adversarial, and the batch size is set to $c_B = \sqrt{c_{\min}} = 10$. Assume that the original set of actions $\mathcal{K}$ is the power set of $\mathcal{I}$:
\begin{equation*}
    \mathcal{K} = \{\{A, B\}, \{A\}, \{B\}, k_0\},
\end{equation*}
where $\{\emptyset\}$ is replaced by a trivial action $k_0$. 

Assume that there is only a single type of request, and for each action $k \in \mathcal{K} = 2^\mathcal{I}$, for each resource type $X \in k \subseteq \{A, B\}$, we use a unit amount of resource $X$, and get the reward of $|k|$.

Consider an online algorithm ALG and its batching extension, ALG-B. For simplicity, assume that ALG is a greedy algorithm: it chooses an action with the highest rewards, and if there is a tie, ALG picks any highest-rewarding action. 

Upon the arrival of the first request, suppose the exogenous replenishment is \(\zeta_A(1) = 1\) and \(\zeta_B(1) = 10\). Since $\zeta_A(1) < c_B$, ALG-B ignores the replenishment for resource type $A$. However, because $\zeta_B(1) = c_B$, a new batched resource type $B_1$ is introduced with different starting inventory $c_{B_1} = 10$. Recall that the name $B_1$ indicates a batched resource type identical to the original resource type $B$, introduced upon the arrival of the first request. Also, note that if we incorporate $\zeta_A(1)$ as the new resource type, say $A_1$, then the resulting instance will start with the minimum starting inventory of $1$, which is no longer desirable in the large-inventory regime.

ALG-B then runs ALG on the modified instance with the extended resource set $\{A, B, B_1\}$ and chooses an action from the updated set of actions $
    \mathcal{K}(1) = \{\{A, B\}, \{A, B_1\}, \{A\}, \{B\}, \{B_1\}, k_0\}.$ Suppose ALG-B (or ALG) selects action $\{A, B_1\}$. Then, when implemented, the reward and resource consumption of this action are identical to those of action $\{A, B\}$.

Now, upon the second request, assume that we receive additional replenishment $\zeta_A(2) = 10$ and \(\zeta_B(2) = 1\). Now, we introduce the new resource type $A_2$ with starting inventory $c_{A_2} = 11$. The updated set of actions becomes $
    \mathcal{K}(2) = \{\{A, B\}, \{A_2, B\}, \{A, B_1\}, \{A_2, B_1\}, \{A\}, \{A_2\}, \{B\}, \{B_1\}, k_0\}.$
    
We may interpret that the new actions involving resource type $A_2$ ($\{A_2, B\}$, $\{A_2, B_1\},$ and $\{A_2\}$) were also available for the first request, but they yield no reward at that time since the resource $A_2$ was not yet available. ALG-B now proceeds by running ALG on this transformed instance, given that it chose an action $\{A, B_1\}$ for the first request. Assume that it chooses $\{A_2, B_1\}$ this time. Then, again, when implemented, ALG-B implements the action $\{A, B\}$.

\paragraph{Example 2 (Under Stochastic Arrival and Replenishment Processes)} Next, to illustrate how batched replenishment is constructed under a stochastic replenishment process, let's use the same instance from \textit{Example 1} but consider only a single request. Also, assume that both the arrival and replenishment processes are stochastic, with \(d = 2\) and \(M = 10\). This means that each action uses at most two resource types, and each exogenous replenishment is bounded by \(10\). For simplicity, assume that \(\epsilon = \left(\frac{3M}{c_{\min}} \log\left(c_{\min} d\right)\right)^{1/3}\) is $0.03$ (though $0.03$ is a hypothetical value), and the batch size is \(c_B = \epsilon c_{\min} = 3\).

Recall that in \textit{Example 1}, we assumed a single type of request, and in this setting, we assume one arrival with probability one, revealed to the stochastic arrival process in advance. For the stochastic replenishment process, assume that the exogenous replenishment is defined as follows:
\(\zeta_A(1), \zeta_B(1)\overset{\text{i.i.d.}}{\sim} X\), where
\begin{equation*}
X =
\begin{cases}
0 \quad \text{with probability } \frac{1}{2},\\
10 \quad \text{with probability } \frac{1}{2}.
\end{cases}
\end{equation*} Note that we get the very same instance as $\textit{Example 1}$ with probability $\frac{1}{4}$. 

Here, assume that ALG is also the greedy algorithm. In ALG-B, we start by constructing $H_B$. It consists of the resource set $\{A, A_1, B, B_1\}$ with the following set of actions: 
\begin{equation*}
  \mathcal{K}(1) = \{\{A, B\}, \{A_1, B\}, \{A, B_1\}, \{A_1, B_1\}, \{A\}, \{A_1\}, \{B\}, \{B_1\}, k_0\}.  
\end{equation*}
The starting inventory of $A_1$ and $B_1$ will be $5(1-\epsilon)$. 

Assume that ALG on $H_B$ selects action $\{A_1, B_1\}$. However, this action can be implemented only if both $\zeta_A(1) = 10$ and $\zeta_B(1) = 10$, which happens with probability $\frac{1}{4}$. Therefore, ALG-B executes $k_0$ with probability $\frac{3}{4}$ or executes $\{A, B\}$ with probability $\frac{1}{4}$. In either case, ALG on $H_B$ believes that the action $\{A_1, B_1\}$ is implemented.

\subsection{Analysis of the Expected LP}\label{appendix-proof-lp}

We first show that the expected LP gives an upper bound to the offline optimal.

\noindent \textbf{Lemma 4.1} \textit{For any instance $H$, $\text{LP}(H) \geq \text{OPT}(H)$.}

\begin{proof}
    For any realization of $A_{ijkz}(l)$'s and $\zeta_i(j)$'s,
    \begin{equation*}
        \sum\limits_{j \in [l]}\sum\limits_{k \in \mathcal{K}}\sum\limits_{z \in Z} A_{ijkz}(l) X_{jkz} \leq c_i + \sum\limits_{j \in [l]} \zeta_i(j), \quad \forall l \in \mathcal{J}, \forall i \in \mathcal{I}.
    \end{equation*}
    Since \(A_{ijkz}(l)\) and \(X_{jkz}\) are independent, taking the expectation of the above inequality gives that \((\mathbb{E}[X_{jkz}])_{j \in \mathcal{J}, k \in \mathcal{K}, z \in Z}\) is a feasible solution to (LP). Therefore, \(\text{LP}(H) \geq \min\limits_{i \in \mathcal{I}}\left(\sum\limits_{j \in \mathcal{J}}\sum\limits_{k \in \mathcal{K}} \sum\limits_{z \in Z} r_{ijkz} \mathbb{E} [X_{jkz}]\right) = \text{OPT}(H)\) by the definition of $\lambda$ in (LP).
\end{proof}

Next, we show that as the initial inventory \(c_i\) increases, the optimal reward $\text{OPT}(H)$ approaches $\text{LP}(H)$. This result leverages the attenuated independent rounding technique and the well-known concentration bounds. 
We use the following standard Chernoff--Hoeffding bounds for sums of
independent random variables; see Theorem~1.1 of
\citet{dubhashi2009concentration}.

\begin{lemma}[Chernoff--Hoeffding bounds]\label{Chernoff}
Let $X_1,\ldots,X_T$ be independent random variables supported on
$[0,1]$, and let
\[
X=\sum_{t=1}^T X_t.
\]
Then, for any $\delta\in(0,1)$,
\begin{align*}
\mathbb{P}\left(X\geq(1+\delta)\mathbb{E}[X]\right)
&\leq
\exp\left(-\frac{\delta^2}{3}\mathbb{E}[X]\right),\\
\mathbb{P}\left(X\leq(1-\delta)\mathbb{E}[X]\right)
&\leq
\exp\left(-\frac{\delta^2}{3}\mathbb{E}[X]\right).
\end{align*}
\end{lemma}

Note that if $X_1,\ldots,X_T$ are independent random variables supported on $[0,M]$ with $M>0$, applying Lemma~\ref{Chernoff} to the rescaled variables $X_t/M\in[0,1]$ yields $\mathbb{P}(X\leq(1-\delta)\mathbb{E}[X])\leq\exp\bigl(-\frac{\delta^2}{3M}\mathbb{E}[X]\bigr)$.

Now, we are ready to prove Lemma \ref{L2.2}. Recall that \(c_{\min} = \min\limits_{i \in \mathcal{I}} c_i \) is the minimum initial inventory across all resource types and $d$ is the maximum number of resource types affected by any action $k \in \mathcal{K}$. Lemma \ref{L2.2} shows that the gap between $\text{LP}(H)$ and $\text{OPT}(H)$ depends on $c_{\min}$ and $d$. 

\noindent \textbf{Lemma 4.2} \textit{For any instance $H$ with a deterministic replenishment process, \[\text{OPT}(H) \geq \left(1- \mathcal{O}\left(\sqrt{\frac{\log (c_{\min}d)}{c_{\min}}}\right)\right) \text{LP}(H).\]}

\begin{proof}
    Denote $(x_{jkz}^*)_{j \in \mathcal{J}, k \in \mathcal{K}, z \in Z}$ as the optimal solution of (LP). For a fixed $\delta \in (0, 1)$, think of the following randomized algorithm:
    
    \begin{center}
        For each arrival \(j \in \mathcal{J}\) with request type $z \in Z$, serve it with action \(k \in \mathcal{K}\) with probability \(\frac{x_{jkz}^*}{(1+\delta) p_{jz}}\). If no action is sampled or implementing the chosen action \(k\) conflicts with any capacity constraint, skip arrival \(j\).
    \end{center}

    Let \(\bar{X}_{jkz}\) be a random variable that equals 1 when action \(k\) is chosen for request $j$ of type $z$ (i.e., \(\bar{X}_{jkz}\) is 1 with probability \(x_{jkz}^*/(1+\delta)\)), and let \(\hat{X}_{jkz}\) be a random variable that equals 1 when action \(k\) is actually implemented by the algorithm. Let $\mathcal{I}_k$ be the subset of $\mathcal{I}$ that action $k$ may consume. By assumption, $|\mathcal{I}_k| \leq d$. The expected revenue of this randomized algorithm is $\min\limits_{i \in \mathcal{I}} \left(\sum\limits_{j \in \mathcal{J}}\sum\limits_{k \in \mathcal{K}}\sum\limits_{z \in Z} r_{ijkz} \E[\hat{X}_{jkz}]\right)$. Note that \begin{equation*}\label{Union}
        \E[\hat{X}_{jkz}] \geq \E[\bar{X}_{jkz}] \mathbb{P} \left( \sum\limits_{t \in [j]}\sum\limits_{k \in \mathcal{K}}\sum\limits_{z \in Z} A_{it kz}(j) \bar{X}_{t kz} \leq c_i + \sum\limits_{t \in [j]} \zeta_i(t), \quad \forall i \in \mathcal{I}_k\right).
    \end{equation*}
    Since $\E [\bar{X}_{jkz}] = \frac{x_{jkz}^*}{1+\delta}$, it suffices to show that we can bound the above probability.

    Fix $j \in \mathcal{J}$. Using the union bound, 
    \begin{align*}
        & \mathbb{P} \left( \sum\limits_{t \in [j]}\sum\limits_{k \in \mathcal{K}}\sum\limits_{z \in Z} A_{itkz}(j) \bar{X}_{t kz} \leq c_i + \sum\limits_{t \in [j]} \zeta_i(t), \quad \forall i \in \mathcal{I}_k\right) \\
        \geq & 1 - \sum\limits_{i \in \mathcal{I}_k} \mathbb{P} \left( \sum\limits_{t \in [j]}\sum\limits_{k \in \mathcal{K}}\sum\limits_{z \in Z} A_{it kz}(j) \bar{X}_{tkz} > c_i + \sum\limits_{t \in [j]} \zeta_i(t)\right). \\
    \end{align*}
    For each $i \in \mathcal{I}_k$, by the definition of $x_{jkz}^*$ and $\bar{X}_{jkz}$, we have 
    \begin{equation*}
        \E [\sum\limits_{t \in [j]}\sum\limits_{k \in \mathcal{K}}\sum\limits_{z \in Z} A_{itkz}(j) \bar{X}_{tkz}] = \frac{\sum\limits_{t \in [j]}\sum\limits_{k \in \mathcal{K}}\sum\limits_{z \in Z} a_{itkz}(j) x^*_{tkz}}{1+\delta} \leq \frac{c_i + \sum\limits_{t \in [j]} \zeta_i(t)}{1+\delta}.
    \end{equation*}
    Using the result of Lemma $\ref{Chernoff}$, 
     \begin{equation*}
        \mathbb{P} \left( \sum\limits_{t \in [j]}\sum\limits_{k \in \mathcal{K}}\sum\limits_{z \in Z} A_{i t k z}(j) \bar{X}_{tkz} > c_i + \sum\limits_{t \in [j]} \zeta_i(t)\right) \leq \exp\left(-\frac{\delta^2 c_{\min}}{3(1+\delta)}\right).
    \end{equation*}
    Choosing
\[
\delta =
\sqrt{\frac{6\log(c_{\min}d)}{c_{\min}}}
\]
for sufficiently large $c_{\min}$ yields
    \begin{equation*}
        \mathbb{P} \left( \sum\limits_{t \in [j]}\sum\limits_{k \in \mathcal{K}}\sum\limits_{z \in Z} A_{i t k z}(j) \bar{X}_{t kz} > c_i + \sum\limits_{ t \in [j]} \zeta_i(t)\right)  \leq  \frac{1}{c_{\min}d}
    \end{equation*}
    for each $i \in \mathcal{I}_k$.
    Therefore, we have $ \E [\hat{X}_{jkz}] \geq \frac{x_{jkz}^*}{1+\delta}\left(1-\frac{1}{c_{\min}}\right) \geq \left(1 - \mathcal{O}(\delta)\right)x_{jkz}^*$, completing the proof of Lemma \ref{L2.2}.
    
\end{proof}

\end{document}